\documentclass[useAMS,usenatbib]{mn2e}

\usepackage{epsfig}
\newcommand{\pars}{Parsamian~21}

\title[High-resolution polarimetry of \pars{}]{High-resolution
  polarimetry of \pars{}: revealing the structure of an edge-on FU\,Ori
  disc\thanks{The results published in this paper are based on data
    collected at the European Southern Observatory in the frame of the
    programme P073.C-0721(A).}}

\author[\'A.~K\'osp\'al et al.]{\'A. K\'osp\'al$^{1}$\thanks{E-mail:
kospal@konkoly.hu}, P. \'Abrah\'am$^{1}$, D.~Apai$^{2,3}$,
  D. R. Ardila$^{4}$, C. A. Grady$^{5}$, \newauthor Th. Henning$^{6}$,
  A. Juh\'asz$^{6}$, D. W. Miller$^{7}$ and A. Mo\'or$^{1}$\\
$^{1}$Konkoly Observatory of the Hungarian Academy of Sciences,
P.O.~Box 67, H-1525 Budapest, Hungary\\
$^{2}$Steward Observatory, The University of Arizona,
933 N.~Cherry Avenue, Tucson, AZ 85721, USA\\
$^{3}$NASA Astrobiology Institute\\
$^{4}$Spitzer Science Center, California Institute of Technology,
Pasadena, CA 91125, USA\\
$^{5}$Eureka Scientific and Goddard Space Flight Center, Code
667, Greenbelt, MD 20771, USA\\
$^{6}$Max-Planck-Institute for Astronomy, K\"onigstuhl 17, 69117
Heidelberg, Germany\\
$^{7}$Department of Physics and Astronomy, University of Louisville,
Louisville, KY 40292, USA\\
}

\begin{document}

\date{Accepted date. Received date; in original form date}

\pagerange{\pageref{firstpage}--\pageref{lastpage}} \pubyear{2007}

\maketitle

\label{firstpage}

\begin{abstract}
  We present the first high spatial resolution near-infrared direct
  and polarimetric observations of \pars{}, obtained with the VLT/NACO
  instrument. We complemented these measurements with archival
  infrared observations, such as HST/WFPC2 imaging, HST/NICMOS
  polarimetry, Spitzer IRAC and MIPS photometry, Spitzer IRS
  spectroscopy as well as ISO photometry. Our main conclusions are the
  following: (1) we argue that \pars{} is probably an FU\,Orionis-type
  object; (2) \pars{} is not associated with any rich cluster of young
  stars; (3) our measurements reveal a circumstellar envelope, a polar
  cavity and an edge-on disc; the disc seems to be geometrically flat
  and extends from approximately 48 to 360 AU from the star; (4) the
  SED can be reproduced with a simple model of a circumstellar disc
  and an envelope; (5) within the framework of an evolutionary
  sequence of FUors proposed by \citet{green} and \citet{quanz},
  \pars{} can be classified as an intermediate-aged object.
\end{abstract}


% --------------------------------------------------------------------
% INTRODUCTION 
% --------------------------------------------------------------------

\section{Introduction}

The most important process of low-mass star formation is the accretion
of circumstellar material onto the young star. According to current
models, the mass accumulation is time-dependent with alternating
episodes of high and low accretion rates \citep{vorobyov,boley}. The
high phase may correspond to FU\,Orionis objects (FUors), a unique
class of low-mass pre-main sequence stars that have undergone a major
outburst in optical light of $4\,$mag or more \citep{herbig77}.
Currently about 20 objects have been classified as FU\,Ori-type
\citep[for a list see][]{fuors}. Records of the outbursts are not
available for all of these objects: some were classified as FUors
since they share many spectral properties with the FUor
prototypes. According to the most widely accepted model \citep{kh91},
the observed spectral energy distribution (SED) is reproduced by a
combination of an accretion disc and an envelope. The energy of the
outburst originates from the dramatic increase of the accretion
rate. On the other hand, \citet{herbig2003} favour a model in which
the outburst occurs in an unstable young star rotating near breakup
velocity.

\pars{} is a system consisting of a central object, HBC\,687
($\alpha_{2000}$ = 19$^{\rm h}$ 29$^{\rm m}$ 0\fs87, $\delta_{2000}$ =
9\degr{} 38\arcmin{} 42\farcs{}69), and an extended nebula first
listed in the catalogue of \citet{par}. Though no optical outburst was
ever observed, \pars{} was classified as a FUor on the basis of its
optical spectroscopic and far-infrared properties \citep{sn92}.
According to \citet{henning} \pars{} is situated in a molecular cloud
named ``Cloud\,A'', very close to the Galactic plane. Indeed, the
distance of 400\,pc and the radial velocity of $v_{lsr} = +27 \pm
15$\,km s$^{-1}$ \citep{sn92} agree very well with the respective
values for Cloud\,A \citep[$v_{lsr} = +27 \pm 4$\,km s$^{-1}$, $d =
500 \pm 100$\,pc,][]{dame85}.  Recently \citet{quanz} questioned the
FUor nature of \pars{} referring to mid-infrared spectral properties,
which better resemble those of post-AGB stars (this issue will be
discussed in Sect.~\ref{sec:discussion}).

The elongated nebulosity around the central source extends
${\approx}\,1\,$arcmin to the north, while it is less developed to the
south. It is visible in the optical and the near-infrared, though the
source becomes unresolved at $10.8$ and $18.2\,\mu$m
\citep{polom}. Polarimetric observations by \citet{draper} revealed a
centrosymmetric polarization pattern and an elongated band of low
polarization perpendicular to the axis of the bright
nebula. \citet{bm90} interpreted the polarization map of \pars{} in
terms of multiple scattering in flattened, optically thick structures
and derived an inclination angle of 80-85\degr{} and a size of 30
$\times$ 8 arcsec for this disc-like structure. The existence of a
disc is also supported by the discovery of a short bipolar outflow
oriented along the polar axis of the nebula \citep{sn92}. Emission at
submm wavelengths was detected by \citet{henning} and
\citet{polom}. The surroundings of \pars{} have been searched for
companions several times, but close-by stars seen in J, H and K-band
images proved to be field stars \citep{li}, and no close-by sources
have been found at longer wavelengths so far \citep{polom}.

Recently several studies were published on the circumstellar
environment of FUors using high-resolution infrared and
interferometric techniques (e.g.~\citealt{green, malbet1998,
malbet2005, millan-gabet, quanz}). Such observations could prove or
disprove the basic assumptions of FUor models. In this paper we
present the highest resolution imaging and polarimetric data yet on
\pars{} from the VLT/NACO instrument, allowing the inspection of the
circumstellar material and disc at spatial scales of ${\approx}\,0.07$
arcsec.  The edge-on geometry of \pars{} represent an ideal
configuration to separate the components of the circumstellar
environment (disc, envelope) and understand their role in the FUor
phenomenon.

% ---------------------------------------------------------------------
% OBSERVATIONS AND DATA REDUCTION
% ---------------------------------------------------------------------

\section{Observations and data reduction}

The log of observations presented in this paper is summarized in
Table~\ref{tab:log}. In the following subsections we describe these
measurements in detail.

% =====================================================================
\subsection{VLT/NACO observations}
\label{sec:naco}

We have acquired ground-based near-infrared imaging and polarimetric
observations in visitor mode using the NACO instrument mounted on the
UT4 of ESO's Very Large Telescope VLT at Cerro Paranal, Chile on 2004
June 18. NACO consists of the NAOS adaptive optics system and the
CONICA near-infrared camera \citep{1998SPIE.3354..606L,
2000SPIE.4008..830H, 2003SPIE.4839..140R}. The weather conditions were
excellent, the typical optical seeing was $\approx 0.72$ arcsec with
two short peaks of $1.3$ arcsec during the night. For imaging we used
the visual dichroic and a camera with 13 mas/pixel scale to obtain H
($\lambda_{\mathrm{c}} = 1.66\,\mu$m) and K$_S$ ($\lambda_{\mathrm{c}}
= 2.18\,\mu$m) images and a camera with 27 mas/pixel scale to obtain
L$^{\prime}$ ($\lambda_{\mathrm{c}} = 3.8\,\mu$m) images. In each
filter a 4-point dithering was applied with small-amplitude random
jitter at each location. The adaptive optics configuration was
fine-tuned for each filter to ensure the best possible correction. The
H-band polarimetric observations were obtained using the 27 mas/pixel
scale camera.

\begin{table*}
\centering          
\caption{Log of the observations of \pars{}. Top part: data from this
  paper; Bottom part: archival data.}
\label{tab:log}      
\begin{tabular}{@{}l@{}c@{}c@{}c@{}c@{}c@{}cc@{}}
\hline
Instrument   & Filter (central $\lambda$)          & Mode         & Date      & Exp. Time                      & Field of View                  & Pixel scale & FWHM      \\
             &                                     &              & YY/MM/DD  &                                & [arcsec $\times$ arcsec]       & [arcsec]    & [arcsec]  \\
\hline                    
VLT/NACO     & H ($1.66\,\mu$m)                    & Imaging      & 04/06/18  & 8$\times$30 s                  & 13.6$\times$13.6               & 0.013       & 0.07      \\
VLT/NACO     & K$_S$ ($2.18\,\mu$m)                & Imaging      & 04/06/18  & 8$\times$20 s                  & 13.6$\times$13.6               & 0.013       & 0.07      \\
VLT/NACO     & L$^{\prime}$ ($3.80\,\mu$m)         & Imaging      & 04/06/18  & 48$\times$0.2 s                & 27.8$\times$27.8               & 0.027       & 0.11      \\
VLT/NACO     & H (1.66 $\mu$m)                     & Polarimetry  & 04/06/18  & 72$\times$10 s, 24$\times$80 s & 27.8$\times$3.1                & 0.027       & 0.12      \\
&&&&&&&\\
GFP          & 6563 and 6590\AA                    & Imaging      & 07/05/15  & 900 s                          & 130 circular                   & 0.38        & 1.1       \\
\hline
HST/NICMOS   & $2\,\mu$m                           & Polarimetry  & 97/11/12  & 18$\times$31 s                 & 19.5$\times$19.3               & 0.076       & 0.16      \\
HST/WFPC2    & F814W ($0.80\,\mu$m)                & Imaging      & 01/07/30  & 2$\times$500 s                 & 79.5$\times$79.5               & 0.1         & 0.17      \\
&&&&&&&\\
Spitzer/IRAC & $3.6$, $4.5$, $5.8$ and $8.0\,\mu$m & Imaging      & 04/04/21  & 848 s                          & 318$\times$318                 & 1.2         & 1.9--2.7  \\
Spitzer/MIPS & $24$ and $70\,\mu$m                 & Imaging      & 04/04/13  & 73 s, 462 s                    & 444$\times$480, 174$\times$192 & 2.5, 4.0    & 6, 18     \\
Spitzer/IRS  & $5-40\,\mu$m                        & Spectroscopy & 04/04/18  & 1121 s                         &                                &             &           \\
&&&&&&&\\
ISO/ISOPHOT  & $65$ and $100\,\mu$m                & Imaging      & 96/09/28  & 3542 s                         & 360$\times$420                 & 15          & 44, 47    \\
\hline                  
\end{tabular}
\end{table*}

% =====================================================================
\paragraph*{VLT/NACO imaging.} The data reduction was carried out
using self-developed IDL routines. For the H and K$_S$ band filters
lampflats were taken, while for the L$^{\prime}$ filter, skyflats were
acquired. For each filter, these images were combined to a final
flatfield, and bad pixel maps were produced. Then the raw images were
flatfield corrected, and bad pixels were removed. Sky frames were
calculated by taking the median of all images taken with the same
filter, then images were sky subtracted. Individual frames were
shifted to the same position and their median was taken to obtain the
final mosaic. Shifts were calculated by computing the
cross-correlation of the images, which gives a more precise result in
case of extended sources, than a simple Gaussian-fitting to the
peak. Dithering resulted in a final mosaic of $21.4\,{\times}\,21.4$
arcsec in case of H and K$_S$ band, and $43.8\,{\times}\,43.8$ arcsec
in L$^{\prime}$ band. The photometric standard was S889-E for the H
and K$_S$ filters (H$\,{=}\,11.662\,{\pm}\,0.004$ mag,
K$_S\,{=}\,11.585\,{\pm}\,0.005$ mag, \citealt{persson}), and
HD\,205772 for the L$^{\prime}$ filter
(L$^{\prime}\,{=}\,7.636\,{\pm}\,0.027$ mag, \citealt{bouchet}).

% =====================================================================
\paragraph*{VLT/NACO polarimetry.} Polarimetric observations were
obtained using the differential polarimetric imaging technique
\citep[DPI, see e.g.][]{draper, kuhn, twhya}. The basic idea of this
method is to take the difference of two orthogonally polarized,
simultaneously acquired images of the same object in order to remove
all non-polarized light. As the non-polarized light mainly comes from
the central star, after subtraction only the polarized light, such as
the scattered light from the circumstellar material remains. This
standard way of dual-beam imaging polarimetry combined with adaptive
optics at the VLT is a powerful way of doing high-contrast mapping.
We obtained polarimetric images with NACO through the H filter, using
a Wollaston prism with a 2 arcsec Wollaston mask to exclude
overlapping beams of orthogonal polarization. \pars{} was observed at
four different rotator angles of 0$^{\circ}$, 45$^{\circ}$,
90$^{\circ}$ and 135$^{\circ}$, providing a redundant sampling of the
polarization vectors. At each angle a 3-point dithering was
applied. The polarimetric calibrator was R CrA DC No.~71
\citep{whittet}. The polarimetric data reduction was done in IDL using
previously developed software tools which we presented in detail in
\citet{twhya}. According to the NACO User's Manual, instrumental
polarization is generally about 2\%. Based on our polarimetric
calibration measurements, we found a typical instrumental polarization
value of 3\% (position angle: 91$^{\circ}$ east of north). Considering
the expected high polarization for the \pars{} nebula, we can conclude
that instrumental polarization can be neglected in the present case,
except in limited areas of low polarization.

% =====================================================================
\subsection{GFP H$_{\alpha}$ imagery}

\pars{} was observed with the Goddard Fabry-Perot (GFP) interferometer
at the Apache Point Observatory 3.5m telescope on 2007 May 15. The
observations were made in direct imaging mode. The on-band image has a
central wavelength of 6563\,\AA{} (H$_{\alpha}$) and a width of 120
km s$^{-1}$. The off-band image has a central wavelength of 6590\,\AA{} and a
width of 120 km s$^{-1}$. The pixel scale is 0.38 arcsec per pixel. The
observations were made through patchy cirrus and are not photometric,
with seeing near 1 arcsec. The instrument and data reduction are
described in \citet{wassell} and references therein. For the
H$_{\alpha}$ on-band image, conspicuous night sky rings contaminate
the region near Par 21. In order to remove these rings, we created a
data cube consisting of subimages centred on the GFP optical axis,
rotated by different angles. After excluding the area around \pars{},
we azimuthally medianed the subimages, and subtracted the resulting
image from the original measurement. For the off-band image, the night
sky rings had displaced beyond the location of the Par 21 nebulosity,
thus no ring subtraction was necessary. The on- and off-band images
were shifted and scaled to match each other by measuring the positions
and fluxes of seven stars in the vicinity of \pars{}. The
continuum-subtracted H$_{\alpha}$ image can be seen in
Fig.~\ref{fig:halpha}.

% =====================================================================
\subsection{HST archival data}

\paragraph*{HST/NICMOS} polarimetric observations were obtained
on 1997 November 12 using the NIC2 camera in MULTIACCUM mode, and the
POL0L, POL120L and POL240L polarizing filters. These filters have a
bandpass between $1.9$ and $2.1\,\mu$m. Raw data files were calibrated
at the STScI with the {\it calnica} v.4.1.1 pipeline. Since these data
have not been published yet, we downloaded the pipeline-processed
files and we did further processing using the IDL-based {\it Polarizer
Data Analysis Software}, which is available through the STScI website
and is described by \citet{mazzuca}. This software package combines
the images obtained through the three polarizers using an algorithm
described in \citet{hines}. We used coefficients appropriate for the
pre-NCS measurements from \citet{hines2}. A 6-point dither pattern was
applied, covering in all about $24\,{\times}\,34$ arcsec. The central
star and the inner parts of the \pars{} nebula can only be seen in two
of the six images, which we shifted and co-added.

\paragraph*{HST/WFPC2.} \pars{} was observed with the WFPC2 on 2001
July 30 through the F814W filter. \pars{} is in the middle of the
image on the WF3 chip (Fig.~\ref{fig:hst}), which has a pixel scale of
0.1 arcsec. Since these data have not been published yet, we used the
High-Level Science Product with the association name `U6FC2801B',
available at the HST archive.

% =====================================================================
\subsection{Spitzer archival data}

Observations of \pars{} were obtained on 2004 April 13 (MIPS), April
18 (IRS) and April 21 (IRAC). The MIPS and IRAC, as well as part of
the IRS data are still unpublished. The IRAC images cover an area of
$5.3\,{\times}\,5.3$ arcmin centred on \pars{}. The
`high-dynamic-range' mode was used to obtain 15 frames in each
position, 5 with 0.4\,s exposure time and 10 with 10.4\,s. The frames
were processed with the SSC IRAC Pipeline v14.0 to Basic Calibrated
Data (BCD) level. The BCD frames were then processed using the IRAC
artifact mitigation software, and finally the artifact-corrected
frames were mosaicked with MOPEX. Photometry for \pars{} was done in
IDL. We used the short exposure images, because \pars{} did not
saturate these frames. We used an aperture of $6$ arcsec and a sky
annulus between $6$ and $12$ arcsec. The aperture corrections were
$1.061$, $1.064$, $1.067$ and $1.089$ for the four channels,
respectively\footnote{IRAC Data Handbook, version 3.0, available at
http://ssc.spitzer.caltech.edu/irac/dh/iracdatahandbook3.0.pdf}. Colour
correction was applied by convolving the observed SED with the IRAC
filter profiles in an iterative way. The resulting colour-corrected
fluxes can be seen in Table \ref{tab:cal}.

Photometry for field stars was obtained in the long exposure images,
because most field stars are fainter than \pars{} and did not saturate
even the long exposure frames. For this purpose we used {\it
  StarFinder}, which is an IDL-GUI based program for crowded stellar
fields analysis \citep{starfinder}. PSF-photometry was calculated for
all stars having S/N ${>}\,3$ and aperture corrections corresponding
to ``infinite'' aperture radius were applied ($0.944$, $0.937$,
$0.772$ and $0.737$ for the four channels, respectively$^1$). Then, we
cross-identified the sources in the four bands and found that 100
field stars are present in all four IRAC images. We plotted these
sources as well as
\pars{} on a [3.6]$-$[4.5] vs.~[5.8]$-$[8.0] colour-colour diagram in
Fig.~\ref{fig:ccirac}. This plot will be discussed in
Sect.~\ref{sec:isolated}.

The MIPS $24$ and $70\,\mu$m images were taken in `compact source
super resolution' mode. At $24\,\mu$m 14 frames (each with 2.62 s
exposure time), at $70\,\mu$m 44 frames (each with 10.49 s exposure
time) were combined into a final mosaic using MOPEX. At $24\,\mu$m
\pars{} saturated the detector, thus we used a model PSF to determine
which pixels are still in the linear regime. Then we fitted the PSF
only to these pixels. The $70\,\mu$m image is not saturated, therefore
we simply calculated aperture photometry in IDL using an aperture
radius of $16$ arcsec, sky annulus between $39$ and $65$ arcsec, and
an aperture correction of 1.741\footnote{MIPS Data Handbook, version
3.2, available at
http://ssc.spitzer.caltech.edu/mips/dh/mipsdatahandbook3.2.pdf}. We
found that the source was point-like at both $24$ and
$70\,\mu$m. Colour correction was applied by convolving the observed
SED with the MIPS filter profiles in an iterative way. The resulting
colour-corrected fluxes can be seen in Table \ref{tab:cal}.

The IRS spectroscopy was carried out in `staring mode' using the Short
Low, the Short High and the Long High modules. Measurements were taken
at two different nod positions and at each nod position three
exposures were taken. We started from BCD level and extracted spectra
from the 2D dispersed images using the Spitzer IRS Custom Extraction
software (SPICE). In the case of the Short Low channel we extracted a
spectrum from a wavelength dependent, tapered aperture around the
star, and also from a sky position. In the case of the high-resolution
channels, we extracted spectra from the full slit. For the Short High
channel, the target was off-slit and most of the stellar PSF fell
outside of the slit. We corrected this spectrum for flux loss using
the measured IRS beam profiles (for more details, see the
Appendix). The resulting complete $5-35\,\mu$m spectrum can be seen in
Figs.~\ref{fig:sed} and \ref{fig:pah}.

% =====================================================================
\subsection{ISO archival data}

ISO measurements of \pars{} at $65$ and $100\,\mu$m were obtained in
'oversampled map' (PHT32) mode on 1996 September 28. These
observations were reduced using a dedicated software package
(P32TOOLS) developed by \citet{pht32}. This tool provides adequate
correction for transients in PHT32 measurements. Absolute calibration
was done by comparing the source flux with the on-board fine
calibration source. In order to extract the flux of our target, a
point spread function centred on \pars{} was fitted to the brightness
distribution on each map. At $65\,\mu$m the PSF could be well fitted
by the ISOPHOT theoretical footprint function. At $100\,\mu$m,
however, the source turned out to be extended. In this case the
brightness distribution was modelled as the convolution of the
standard theoretical footprint function with a two-dimensional
elliptical Gaussian of 40 arcsec $\times$ 20 arcsec, and a position
angle of $114^{\circ}$. The obtained fluxes were colour corrected by
convolving the observed SED with the ISOPHOT filter profiles in an
iterative way. The results are displayed in Table~\ref{tab:cal}.

\begin{table}
\caption{Photometry for \pars{}. All fluxes are colour-corrected.}
\label{tab:cal}
\centering
\begin{tabular}{ccc@{}c@{}c}
\hline
Instrument    & Wavelength [$\mu$m] & \multicolumn{3}{c}{Flux [mJy]} \\
\hline
Spitzer/IRAC  & 3.6                 & 143     & $\,\pm\,$ & 8            \\
Spitzer/IRAC  & 4.5                 & 187     & $\,\pm\,$ & 10           \\
Spitzer/IRAC  & 5.8                 & 303     & $\,\pm\,$ & 16           \\
Spitzer/IRAC  & 8.0                 & 829     & $\,\pm\,$ & 42           \\
&&\\
Spitzer/MIPS  & 24                  & 5\,530  & $\,\pm\,$ & 310          \\
Spitzer/MIPS  & 70                  & 13\,300 & $\,\pm\,$ & 1\,300       \\
&&\\
ISO/ISOPHOT   & 65                  & 12\,600 & $\,\pm\,$ & 1\,300       \\
ISO/ISOPHOT   & 100                 & 14\,200 & $\,\pm\,$ & 1\,900       \\
\hline
\end{tabular}
\end{table}

% ---------------------------------------------------------------------
% RESULTS
% ---------------------------------------------------------------------

\section{Results}

\subsection{Broad-band imaging}
\label{sec:imaging}

\begin{figure}
\centering
\includegraphics[angle=0,width=0.95\linewidth]{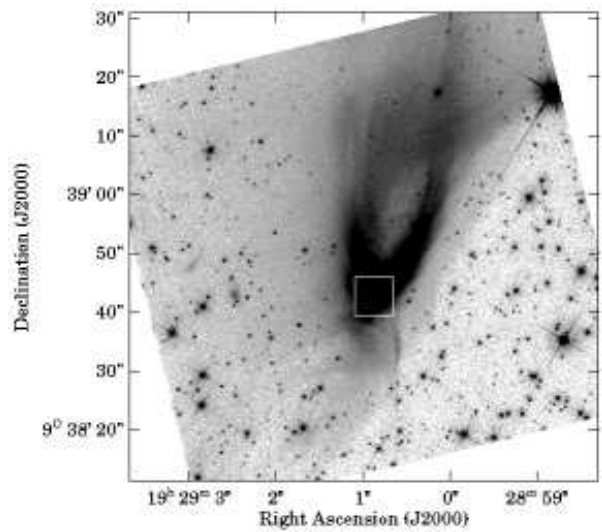}
\caption{HST/WFPC2 image of \pars{} taken through the F814W
  filter. The intensity scale is square root and brightness increases
  from light to dark. The white square marks the area shown in the
  upper left panel of Fig.~\ref{fig:images}.}
\label{fig:hst}
\end{figure}

\begin{figure}
\centering
\includegraphics[angle=0,width=0.95\linewidth]{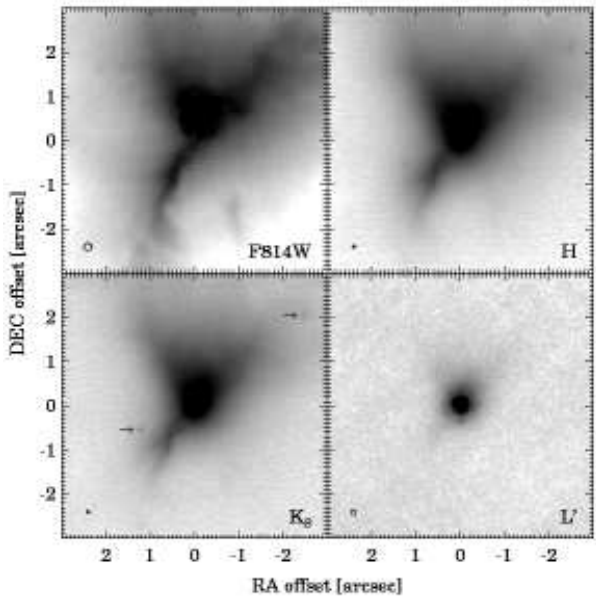}
\caption{HST/WFPC2 (with filter F814W) and VLT/NACO (with filters H,
  K$_S$ and L$^{\prime}$) images of \pars{}. Circles at the left
  bottom corners indicate the FWHM at the centre of the corresponding
  image. In the K$_S$-band image, arrows mark the positions of the two
  closest stars to \pars{} (see Sect.~\ref{sec:isolated}).}
\label{fig:images}
\end{figure}

Figure~\ref{fig:hst} displays the HST/WFPC2 image taken at
$0.8\,\mu$m. The image reveals the structure of the \pars{} nebula
with an unprecedented spatial resolution and detail. We note that
strong emission lines associated with the bipolar outflow (see
Sect.~\ref{sec:halpha}) are not included in the bandpass, so the HST
image provides a clean picture of the reflection nebula. The overall
dimensions of the nebula are approximately $60\,{\times}\,20$
arcsec. The northern part has a characteristic elliptic, loop-like
shape, with the inside of the loop having a relatively low surface
brightness. The outer edge of the nebula is rather well-defined, while
the inner edge is more fuzzy. Apart from the loop itself, there is
also a wide, faint SE-NW oriented stripe about $30$ arcsec from the
star. The nebula is clearly bipolar, although very asymmetric: the
southern part is much less developed than the northern one. We note
that while the southeastern arc is brighter close to the star (see the
central part of the HST image in Fig.~\ref{fig:images}), the
southwestern arc is more conspicuous farther from the star
(Fig.~\ref{fig:hst}). It is remarkable that field stars are visible
both to the north and to the south of \pars{}. Supposing that these
stars are in the background, this implies that there is no significant
difference in the extinction between the northern and southern part.

Figure~\ref{fig:images} shows the central parts of the HST/WFPC2 and
the VLT/NACO images (note that the innermost $0.57$ arcsec of the HST
image is saturated). The bipolar nature of the nebula is most
conspicuous at the shortest wavelength ($0.8\,\mu$m), where all 4 arcs
(roughly to the northeast, northwest, southwest and southeast) are
visible. In the K$_S$ and H bands the nebula is more
triangle-shaped. In all three NACO bands the central source is
extended, with deconvolved sizes of $0.20$, $0.12$ and $0.08$ arcsec
in the H, K$_S$ and L$^{\prime}$ bands, respectively.

\subsection{Narrow-band imaging}
\label{sec:halpha}

\begin{figure}
\centering
\includegraphics[angle=90,width=0.9\linewidth]{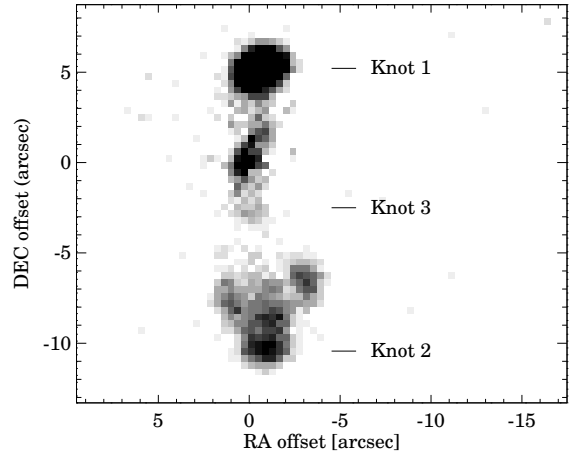}
\caption{Continuum-subtracted H$_{\alpha}$ image of \pars{}. The peak
at position (0,0) is the residuum of the central source. Three
Herbig-Haro knots could be identified; these were named Knot 1, 2 and
3 by \citet{sn92}.}
\label{fig:halpha}
\end{figure}

\begin{figure}
\centering
\includegraphics[angle=90,width=0.9\linewidth]{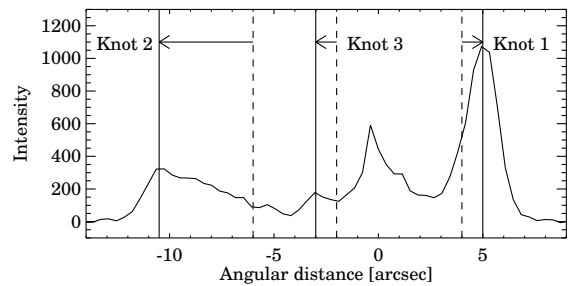}
\caption{South-north oriented cut across \pars{} in the
  continuum-subtracted H$_{\alpha}$ image. 0 marks the position of the
  central star. The peaks marked by solid vertical lines at $-10.5$,
  $-3$ and $5$ arcsec correspond to Knot 2, 3 and 1,
  respectively. Dashed vertical lines mark the positions of the same
  knots as measured by \citet{sn92}. Arrows demonstrate the direction
  where the knots propagate.}
\label{fig:diff_cut}
\end{figure}

Figure~\ref{fig:halpha} displays the continuum-subtracted H$_{\alpha}$
image of \pars{}.  Conspicuous Herbig-Haro knots are seen to the north
and the south of the central star. The northern knot is compact,
resembling the knot described in \citet{sn92}. The southern knot,
however, has a distinctly bowed shape. The southern extent of this
knot has a S/N of 10 over background per pixel, while the flanking
structures have S/N $\approx$ 7. The northern knot has S/N in excess
of 15 per pixel. To measure the angular distance of the knots from the
central star in the H$_{\alpha}$ imagery, we binned the data in a 3
pixel (1.1 arcsec) wide swath oriented north-south. Apart from the
residual of the central source, we could identify three knots (see
Fig.~\ref{fig:halpha}): Knots 1, 2 and 3 are located 5 arcsec north,
10.5 arcsec south and 3 arcsec south, respectively. The knots
correspond to those identified by \citet{sn92}, but all three knots
have moved outwards since then (Fig.~\ref{fig:diff_cut}). Assuming a
distance of $400\,$pc, and supposing that the projected velocities are
close to the velocities of the knots along the polar axis, one can
derive $120\,$km s$^{-1}$ for Knots 1 and 3, and $530\,$km s$^{-1}$
for Knot 2. This is in good agreement with the velocities calculated
by \citet{sn92} using radial velocity measurements. Such outflow
velocities imply kinematic ages of approximately $80\,$yr for Knot 1
and $40\,$yr for Knots 2 and 3.

\subsection{Spectral energy distribution}
\label{sec:sed}

\begin{figure}
\centering
\includegraphics[angle=0,width=0.95\linewidth]{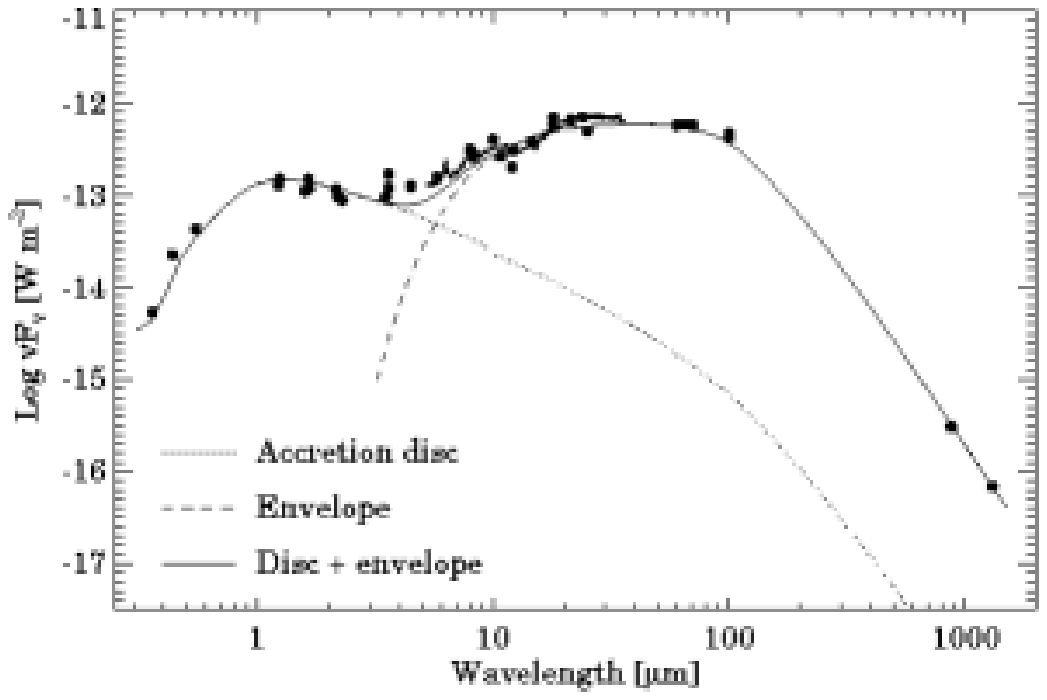}
\caption{Complete UV-to-mm SED of \pars{}. Source of data: 2MASS
  All-Sky Catalog of Point Sources, MSX6C Infrared Point Source
  Catalog, \citet{fuors}, \citet{polom}, \citet{ns84}, \citet{henning}
  and this work. The model SED is discussed in
  Sect.~\ref{sec:modelling}.}
\label{fig:sed}
\end{figure}

\begin{figure}
\centering
\includegraphics[angle=90,width=0.95\linewidth]{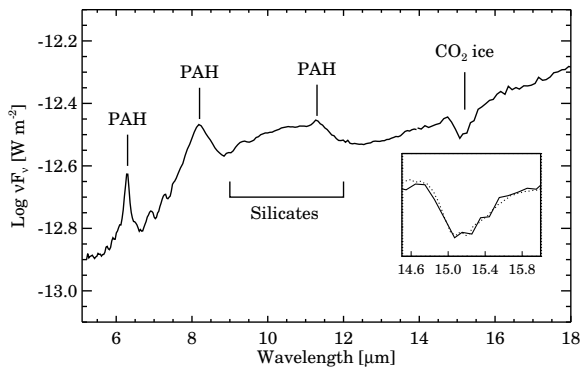}
\caption{$5\,{-}\,18\,\mu$m part of the Spitzer/IRS spectrum of
  \pars{}. The inset displays in details the continuum-subtracted
  spectral region around the $15.2\,\mu$m CO$_2$ ice feature of
  \pars{} with solid line and as a comparison HH\,46\,IRS with dashed
  line (for discussion see Sect.~\ref{sec:general}).}
\label{fig:pah}
\end{figure}

Table~\ref{tab:cal} contains our Spitzer and ISO photometry for
\pars{}. These data, complemented with previously published
measurements between 1983 and 1999 are plotted in Fig.~\ref{fig:sed},
showing the complete UV-to-mm spectral energy distribution (SED) of
the object. Comparing data taken between 1983 and 1999, \citet{fuors}
found that there is no long-term flux variation in the $1-100\,\mu$m
wavelength range. The new Spitzer data from 2004 reveal that the
mid-infrared brightness of \pars{} stayed constant since then at a
$15\%$ level. Although \citet{par96} reported $2-3$ mag brightness
variations in the B band between 1966 and 1990 (in Fig.~\ref{fig:sed}
we plotted optical measurements from 1980 as representative values),
no significant flux changes can be seen at longer wavelengths within
the measurement uncertainties.

At wavelengths shorter than $3\,\mu$m, the SED is highly reddened
\citep{polom}, probably due to the combined effect of interstellar
extinction and self-shadowing by circumstellar material. The fact that
the central source is extended at these wavelengths (see
Sect.~\ref{sec:imaging}) indicates that the photometry is contaminated
by scattered light from the inner part of the circumstellar
environment. Between $3$ and $25\,\mu$m, the SED is rising towards
longer wavelengths as $\nu F_{\nu}\,{\propto}\,\lambda$. The
$5\,{-}\,18\,\mu$m Spitzer spectrum (Fig.~\ref{fig:pah}) displays
strong PAH emission features at $6.3$, $8.2$ and $11.3\,\mu$m and
amorphous silicate emission around $10\,\mu$m (\citealt{quanz}, see
also the spectrum of \citealt{polom}). Weaker absorption features can
also be seen between $7$ and $8\,\mu$m, which cannot be safely
identified with any ice or PAH feature. The $10\,{-}\,19\,\mu$m
channel of the Spitzer/IRS spectrum was not shown by \citet{quanz} due
to data reduction difficulties. We made an attempt to evaluate also
this channel (for details, see Appendix). As can be seen in the inset
of Fig.~\ref{fig:pah}, the spectrum shows a strong CO$_2$ ice
absorption feature at $15.2\,\mu$m, which has not been reported in the
literature so far. Between $25$ and $100\,\mu$m the SED is flat ($\nu
F_{\nu}\,{\propto}\,$const.), while the submillimetre shape is $\nu
F_{\nu}\,{\propto}\,\lambda^{-3.5}$ (corresponding to
$\beta\,{=}\,0.5$, assuming optically thin emission and a dust opacity
law of $\kappa_{\nu}\,{\propto}\,\nu^{\beta}$). Using a distance of
400\,pc and an interstellar reddening of $A_V\,{=}\,2\,$mag
\citep{hillen}, the bolometric luminosity computed as the integral of
the SED from $0.44$ to $1300\,\mu$m is $10\,$L$_{\odot}$ (assuming
isotropic radiation field).

\subsection{Polarimetry}
\label{sec:pol}

\begin{figure*}
\centering
\includegraphics[angle=90,width=0.95\linewidth]{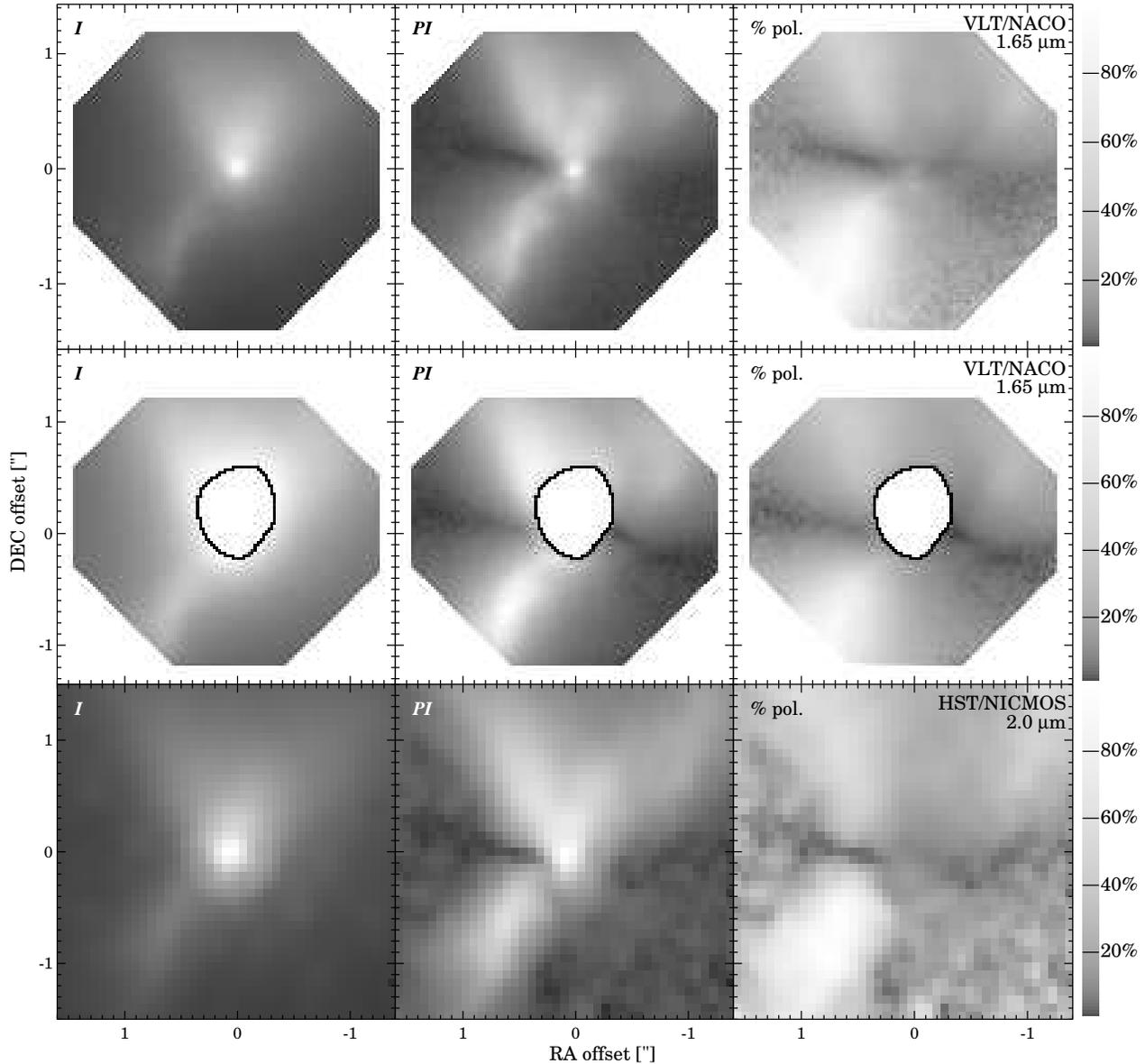}
\caption{{\it Left:} total intensity ($I$); {\it middle:} polarized
  intensity ($PI$); {\it right:} degree of polarization (\% pol.),
  with scalebar. {\it Upper row:} short exposure H-band NACO images;
  {\it middle row:} long exposure H-band NACO images (the saturated
  central parts are masked out); {\it lower row:} $2\,\mu$m NICMOS
  images of the same area. The intensity scale is logarithmic and
  brightness increases from dark to light.}
\label{fig:pol}
\end{figure*}

\begin{figure}
\centering
\includegraphics[angle=90,width=0.95\linewidth]{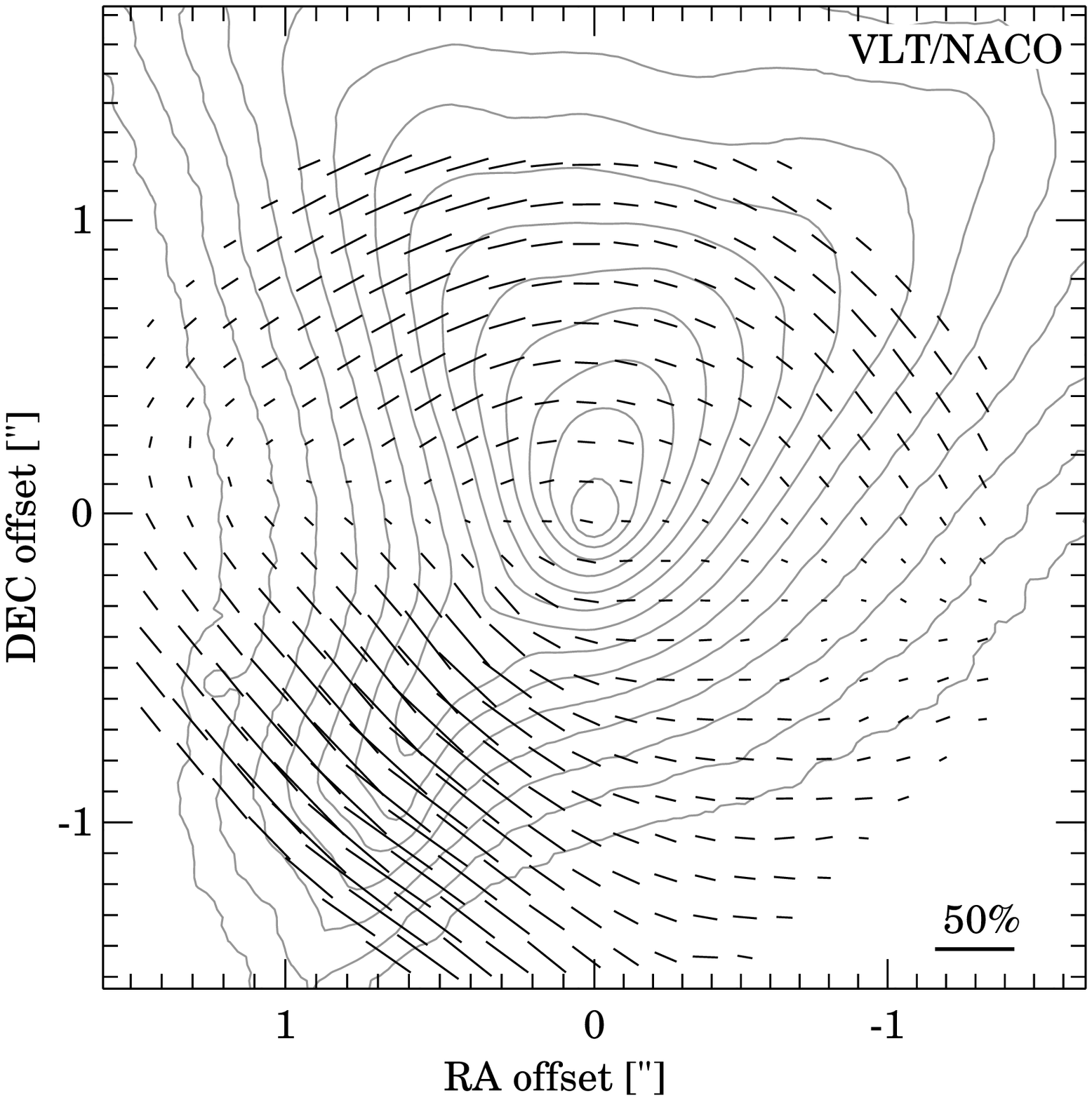}
\includegraphics[angle=90,width=0.95\linewidth]{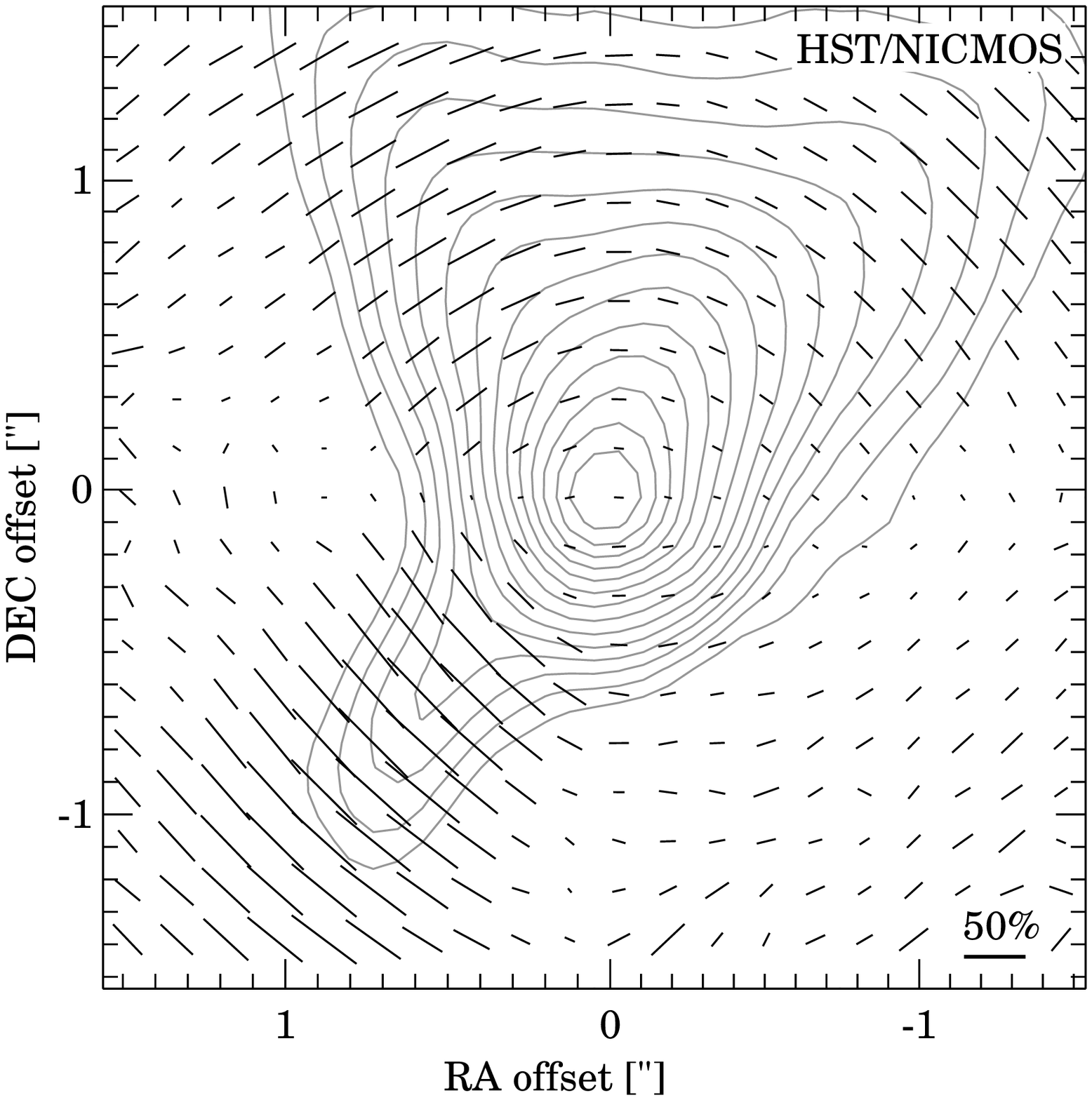}
\caption{Polarization pseudovectors overlaid on intensity
contours. {\it Top:} VLT/NACO, H-band, with $5\,{\times}\,5$ pixels
binning, {\it bottom:} HST/NICMOS, $2\,\mu$m, with $2\,{\times}\,2$
pixels binning.}
\label{fig:vector}
\end{figure}

Figure~\ref{fig:pol} shows images of \pars{} calculated from the
H-band NACO polarization data, as well as from the $2\,\mu$m NICMOS
measurements. The left column shows the total intensity ($I$) obtained
as the sum of the intensities of the two orthogonal polarization
states. The middle column displays the polarized intensity ($PI$)
calculated as the square root of the quadratic sum of the $Q$ and $U$
Stokes components. The right column shows the degree of polarization
(\% pol.), i.e.~the ratio of the polarized to the total intensity.

The NACO total intensity maps (Fig.~\ref{fig:pol}, left) are very
similar to the H-band map in Fig.~\ref{fig:images}. Thus, the
Wollaston-prism data can reproduce well the direct images taken
without the prism. While the total intensity is relatively smooth, the
polarized intensity (Fig.~\ref{fig:pol}, middle) reveals many fine
details not visible in the total intensity map. The most striking
feature is a dark horizontal lane across the star, where the polarized
intensity is very low. Above this dark lane, high intensity regions
can be seen, primarily emphasizing the walls of the upper lobe. Below
the dark lane, the southeastern arc is very pronounced. The degree of
polarization (Fig.~\ref{fig:pol}, right) again shows features
different from the previous two maps. Here, the central star almost
disappears, as the stellar light is not polarized. The horizontal dark
lane of low polarization across the star is even more pronounced:
there the degree of polarization is around ${\approx}\,5\%$ in the
H-band NACO images. We note that the polarization pattern in this area
is probably affected by instrumental polarization
(Sec.~\ref{sec:naco}), though this effect should not alter our
discussion in Sec.~\ref{sec:disc}. The polarization of the central
source itself, measured in a $0.08$ arcsec radius aperture, is $9\%$
and its position angle is $70^{\circ}$ east of north. The long
exposure NACO images (Fig.~\ref{fig:pol}, middle row), showing the
outer regions with higher signal-to-noise ratio, are consistent with
the short exposure maps (Fig.~\ref{fig:pol}, upper row). It is
remarkable that the NICMOS images, which represent independent
observations (different resolution, different instrument, different
polarization technique), show strikingly similar features, although
the degree of polarization is slightly higher.

Figure~\ref{fig:vector} shows the polarization pseudovectors overlaid
on total intensity contours. The NACO map, along with the degree of
polarization map in Fig.~\ref{fig:pol} right shows that the horizontal
band of low polarization across the star is very well-confined, narrow
(about $0.15-0.20$ arcsec wide at the $5\%$ polarization level) and
can be discerned from $0.12$ to $1.2$ arcsec to the star (and likely
beyond, as indicated by the long exposure maps, see also the
polarization maps of \citealt{draper} and \citealt{hajjar}). The
position angle of this band is $78^{\circ}\,{\pm}\,4^{\circ}$ east of
north, consistent with that measured by \citet{draper}
($75^{\circ}\,{\pm}\,4^{\circ}$). This implies that the band in our
high spatial resolution images are a direct inward continuation of the
band seen by \citet{draper} and \citet{hajjar} in their lower
resolution images. The degree of polarization in this band is a few
percent, and it is oriented parallel with the band itself.

Most of the northern part of the nebula shows a centrosymmetric
pattern, characteristic of reflection nebulae with single
scattering. This is consistent with what \citet{draper} and
\citet{hajjar} measured for \pars{} itself. It is also similar to what
can be seen in other bipolar nebulae (see e.g.~\citealt{meakin} for
nebulae of young stars or \citealt{scarrott} for nebulae of evolved
stars). The highest polarization in the northern part can be measured
in the northeastern wall ($25-30\%$) and also in a spot about $0.9$
arcsec west and $0.6$ arcsec north to the star ($20-25\%$). The
southern part of the nebula also follows the regular centrosymmetric
pattern seen in the northern part. Here the highest polarization can
be found in the southeastern arc ($60-70\%$).

The NICMOS vector map at $2\,\mu$m (Fig.~\ref{fig:vector} bottom)
repeats all the main features seen by NACO in the H-band
(Fig.~\ref{fig:vector} top). We note nevertheless the presence of two
small, localized depolarization areas on either side of the central
source at positions ($0\farcs9$, $0\farcs0$) and ($-0\farcs8$,
$-0\farcs1$).

% ---------------------------------------------------------------------
% DISCUSSIONS
% ---------------------------------------------------------------------

\section{Discussion}
\label{sec:discussion}

\subsection{\pars{}: an FU\,Orionis-type star}

Although no optical outburst was ever observed, \pars{} shares many
properties characteristic of FUors. Its spectral type in the
literature ranges from A5 to F8 supergiant \citep[e.g.~][]{sn92},
similar to the typical FUor spectral types (F--G supergiant). In
addition, it shows strong infrared excess and drives a bipolar
outflow, similarly to many FUors. However, in a recent paper by
\citet{quanz} the pre-main sequence nature and the FUor status of
\pars{} were questioned, mainly because its PAH emission bands are
untypical for young stars. They suggest that \pars{} is either an
intermediate mass FUor object, or an evolved star sharing typical
properties with post-AGB stars.

An important parameter which may discriminate between a pre-main
sequence object and a post-AGB star is the luminosity. The typical
luminosity of post-AGB stars is in the order of
$10^3-10^4\,$L$_{\odot}$ \citep{kwok}, while FUors are typically
fainter than a few $100\,$L$_{\odot}$. The luminosity of \pars{},
$10\,$L$_{\odot}$ (see Sect.~\ref{sec:sed}), is at least a factor of
100 lower than that of post-AGB stars. Even adopting the largest
distance estimate mentioned in the literature (1800\,pc,
\citealt{sn92}), the luminosity of \pars{} is still too low. The
observed proper motion of the Herbig-Haro knots, however, prefers the
lower distance (400\,pc), otherwise the outflow velocities
(${\approx}\,2400\,$km s$^{-1}$ at 1800\,pc) would be unusually high
for a young stellar object \citep[typically up to $600\,$km s$^{-1}$,
see e.g.~][]{mundt, hessman}. Moreover, the existence of Herbig-Haro
outflows are usually tracers of low-mass star formation \citep{hh}.

In the following discussion, we consider \pars{} to be a FUor and we
discuss its properties in the context of young eruptive
stars. Nevertheless we note that most of our results on the morphology
and circumstellar structure are valid regardless of the nature of the
central object.

\subsection{\pars{}: an isolated young star?}
\label{sec:isolated}

\begin{figure}
\centering
\includegraphics[angle=0,width=0.95\linewidth]{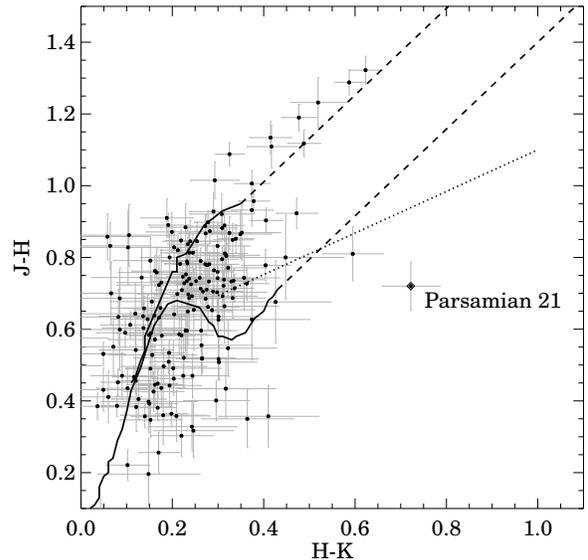}
\caption{2MASS colour-colour diagram of 188 sources in an area of
$5.3\,{\times}\,5.3$ arcmin centred on \pars. The main-sequence and
giant branch are marked by solid lines \citep{koornneef}, the
reddening path with dashed lines \citep{cardelli} and the T\,Tauri
locus with dotted line \citep{meyer}.}
\label{fig:cc2mass}
\end{figure}

\begin{figure}
\centering
\includegraphics[angle=0,width=0.95\linewidth]{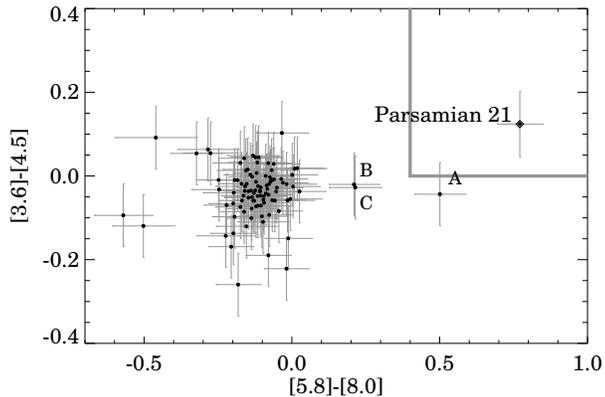}
\caption{IRAC colour-colour diagram of 100 sources located in the same
  area as in Fig.~\ref{fig:cc2mass}. The gray square in the upper
  right corner marks the approximate domain of Class II sources
  \citep{allen}.}
\label{fig:ccirac}
\end{figure}

\begin{table}
\caption{VLT/NACO photometry for the two closest stars. Uncertainties
  are about 0.1 mag.}
\label{tab:twostars} 
\centering          
\begin{tabular}{c c c c c}
\hline
$\alpha_{2000}$                       & $\delta_{2000}$                  & H mag & K$_S$ mag & L$^{\prime}$ mag \\
\hline
19$^{\rm h}$ 29$^{\rm m}$ 0\fs96      & 9\degr{} 38\arcmin{} 42\farcs11  & 20.2  & 19.1      & $>$14.6          \\
19$^{\rm h}$ 29$^{\rm m}$ 0\fs70      & 9\degr{} 38\arcmin{} 44\farcs78  & 20.6  & 19.5      & $>$14.4          \\
\hline                  
\end{tabular}
\end{table}

\pars{} is situated close to the Galactic plane ($l=45.8^{\circ}$,
$b=-3.8^{\circ}$), in a molecular cloud called Cloud\,A. This cloud
was identified by \citet{dame85} in their CO survey of molecular
clouds in the northern Milky Way. The cloud occupies an area of 8
square degrees in the sky, and the only known young star associated
with it is the T\,Tauri star AS\,353 \citep{dame85}.

In order to check whether FUors are usually associated with star
forming regions, we searched the literature and found that most FUors
are located in areas of active star formation
\citep[e.g.][]{henning}. To find out whether there is star formation
in the vicinity of \pars{}, we searched for pre-main sequence
stars. For this purpose we constructed a 2MASS J$-$H vs.~H$-$K$_S$ and
an IRAC [3.6]$-$[4.5] vs.~[5.8]$-$[8.0] colour-colour diagram for
sources found in our $5\farcm3\,{\times}\,5\farcm3$ IRAC field of view
(Fig.~\ref{fig:cc2mass}, \ref{fig:ccirac}). Our selection criteria in
the case of 2MASS was S/N ${>}\,10$ and uncertainties ${<}\,0.1\,$mag
in all J, H and K$_S$ bands, while in the case of IRAC S/N ${>}\,3$
and detectability at all four bands were required.  The 2MASS diagram
revealed that most of the nearby objects are reddened main sequence or
giant stars. On the IRAC diagram, however, there are three objects
(apart from \pars{} itself), which display infrared excess at
$8\,\mu$m (marked by A, B and C in Fig.~\ref{fig:ccirac}). According
to the classification of \citet{allen}, Class II sources exhibit
colours of [3.6]$-$[4.5] ${>}\,0.0$ and [5.8]$-$[8.0] ${>}\,0.4$, thus
one of these stars (A) might be a Class II source, while B and C are
more likely Class III/main sequence sources. The nature of source A
and its possible relationship to \pars{} is yet to be investigated.
Nevertheless, \pars{} seems to be rather isolated compared to most
FUors and certainly not associated with any rich cluster of young
stellar objects.

We also searched for possible close companions of \pars{} in the WFPC2
and NACO direct images. In order to establish a detection limit for
source detection, we measured the sky brightness on the NACO images
(before sky-subtraction), and estimated a limiting magnitude for each
filter. The resulting values are $22.8$, $21.6$, and $15.2\,$mag in H,
K$_S$ and L$^{\prime}$, respectively. In case of the HST/WFPC2 image,
the larger ($80\,{\times}\,80$ arcsec) field of view made it possible
to estimate a limiting magnitude using star counts; the resulting
value is $23.5\,$mag. Due to the bright reflection nebula, the
detection limit is somewhat lower close to the star. The two closest
objects we found are the following: one star to the southeast, at a
distance of $1.4$ arcsec (560\,AU at 400\,pc), and another one to the
northwest, at a distance of $3.3$ arcsec (1320\,AU at 400\,pc). These
sources are marked with arrows in Fig.~\ref{fig:images}. Neither of
the stars are visible at $3.8$ or $0.814\,\mu$m, although we can give
an upper limit for their L$^{\prime}$ brightness. Their positions and
photometry are given in Table~\ref{tab:twostars}. As these sources are
very red, they can equally be heavily reddened background stars, or
stars with infrared excess (indicating that they might be associated
with \pars{}). Supposing that they are reddened main sequence stars,
one can estimate an extinction of
A$_V\,{\approx}\,10\,{-}\,15\,$mag. Further multifilter observations
may help to clarify the nature of these objects and their possible
relationship to \pars{}.

\subsection{The circumstellar environment of \pars{}}

\begin{figure}
%\sidecaption
\centering
\includegraphics[angle=0,width=0.45\linewidth]{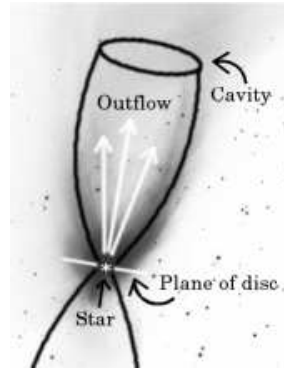}
\caption{Sketch of the morphology of circumstellar material around
\pars{}, overlaid on the HST/WFPC2 image. The central star is
surrounded by an edge-on disc. Perpendicular to the disc, the star
drives a bipolar outflow that excavates an outflow cavity in the dense
circumstellar material. Light from the central star illuminates the
walls of the cavity.}
\label{fig:morph}
\end{figure}

\begin{figure}
\centering
\includegraphics[angle=90,width=0.95\linewidth]{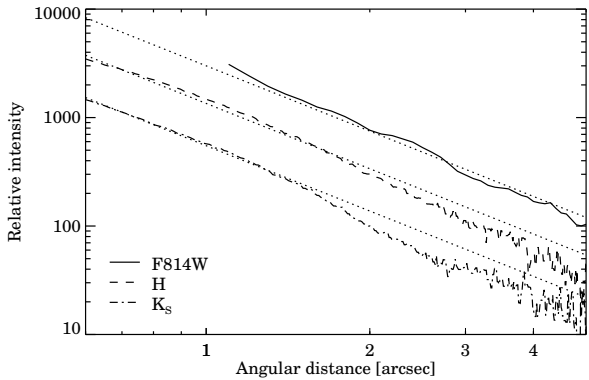}
\caption{Brightness profiles of \pars{} at $0.8\,\mu$m and in the H
and K$_S$ bands. Dotted lines mark Hubble's relation.}
\label{fig:metszet}
\end{figure}

\begin{figure}
\centering
\includegraphics[angle=0,width=0.95\linewidth]{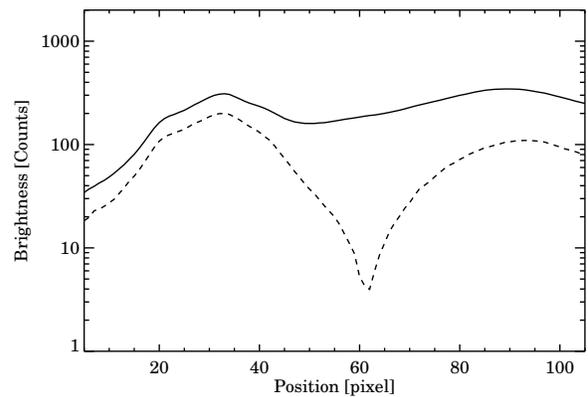}
\caption{South-north cut at $0.6$ arcsec east from the star. {\it
    Solid line:} total intensity; {\it dashed line:} polarized
    intensity.}
\label{fig:cutok}
\end{figure}

\begin{figure*}
\centering
\includegraphics[angle=90,width=0.95\linewidth]{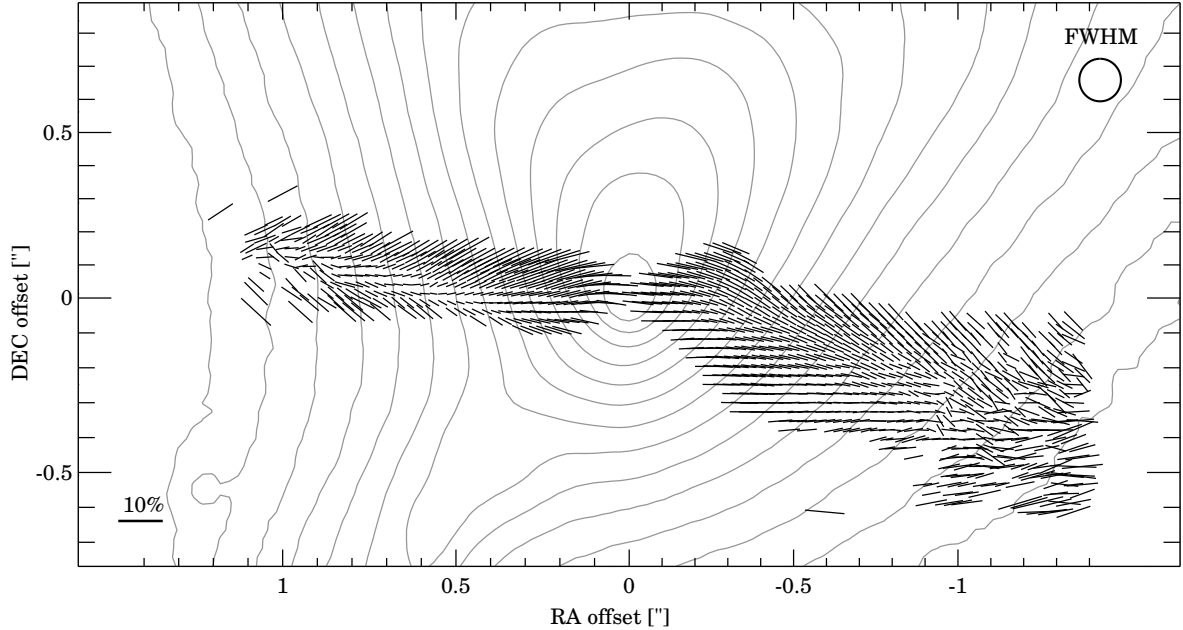}
\caption{VLT/NACO polarization map of \pars{} overlaid on H-band total
  intensity contours. The circle in the upper right corner displays
  the FWHM of the polarization measurement. Polarization vectors are
  displayed at full resolution, only showing the central low
  polarization band, where the polarization vectors are aligned. Such
  arrangement is expected when multiple scattering occurs in an
  edge-on disc.}
\label{fig:kiskep}
\end{figure*}

The appearance and polarization properties of the nebula around
\pars{} can be understood in the following way: the star drives an
approximately north-south oriented bipolar outflow, which had
excavated a conical cavity in the dense circumstellar material
(Fig.~\ref{fig:morph}). The star illuminates this cavity and the light
is scattered towards us mainly from the walls of the cavity. The
outflow direction is perpendicular to an almost edge-on dense
circumstellar disc. This picture is supported by the following facts:
(a) the centrosymmetric polarization pattern is characteristic of
reflection nebulae with single scattering; (b) the morphology and limb
brightening suggest a hollow cavity (as opposed to an ``outflow
nebula'', where the lobes are composed of dense material ejected by
the central source); and (c) the low-polarization lane across the star
strongly suggests the presence of an edge-on circumstellar disc, where
multiple scattering occurs. In general the \pars{} system shows
similarities to the NGC\,2261 nebula associated with R\,Mon. This
object also consists of a northern cometary nebula and a southern
jetlike feature \citep{warren}.

\subsubsection{Envelope/Cavity}
\label{sec:env}

We characterised the opening of the upper lobe by marking the ridge
along the northeastern and northwestern arcs which we interpret as the
walls of the cavity. As viewed from the star northwards, the cavity
starts as a cone with an opening angle of ${\approx}\,60^{\circ}$,
giving the nebula in Fig.~\ref{fig:images} a characteristic
equilateral triangle-shape. Farther away from the star the cavity
deviates from the conical shape, becomes narrower. The whole cavity
occupies an area of $8\,000\,{\times}\,24\,000\,$AU (at a distance of
$400\,$pc). The sharp outer boundary of the nebula implies a
significant density contrast between the cavity and the surrounding
envelope.

In Fig.~\ref{fig:metszet} we plotted radial brightness profiles at
$0.8\,\mu$m and in the H and K$_S$ bands, starting from the star
northwards. The slope of the intensity profiles in the inner
${\sim}\,1.6''$ ($640\,$AU) follows closely Hubble's relation
\citep{hubble}, i.e.~the brightness is proportional to r$^{-2}$. This
trend can be clearly followed above the noise out to about $5\,$arcsec
in our H and K$_S$ images, giving an estimate of $2000\,$AU for the
outer size of the envelope. The fact that the brightness profiles
follow Hubble's relation implies that the nebula is produced by
isotropic single scattering, in accordance with the centrosymmetric
polarization pattern and the high degree of polarization. Around
${\sim}\,1.6''$ the profiles becomes steeper, probably due to
decreased density in the inside of the cavity. Similar steepening in
the outer part of the nebula was already mentioned by \citet{li}. The
profiles are similar at all observed wavelengths and do not show
significant dependence on the position angle.

\subsubsection{Disc}
\label{sec:disc}

As mentioned in Sect.~\ref{sec:pol}, both the NACO and NICMOS
polarimetric images show a lane of low polarization oriented nearly
east-west across the star (Fig.~\ref{fig:pol}). The drop in the degree
of polarization can also be clearly seen in Fig.~\ref{fig:cutok},
where a north-south cut at $0\farcs6$ east to the star is plotted.
Following \citet{bm90}, we interpret the low polarization by multiple
scattering in an edge-on disc (possible other explanations for the
origin of the low polarization areas are discussed in
e.g.~\citealt{lr}). According to models of such circumstellar
structures \citep[e.g.][]{whitney, fischer} the polarization vectors
are oriented parallel to the disc plane. As can be seen in
Fig.~\ref{fig:kiskep}, despite the low degree of polarization, the
predicted alignment of the vectors can be clearly seen in the case of
\pars{} too. The NICMOS polarization map shows a similar effect. The
dark lane in Fig.~\ref{fig:pol} can be followed inwards to as close as
$48\,$AU from the star. This is an upper limit for the inner radius of
the circumstellar disc.

An interesting feature of the mentioned models is two depolarized
areas on either side of the central star, which mark the outer end
points of an edge-on disc. The depolarization is due to a transition
from the linearly aligned to the centro-symmetic polarization
pattern. These depolarized areas can be seen in Fig.~\ref{fig:vector}
bottom, on both sides at about $0.9''$ from the star. At the distance
of \pars{} this corresponds to $360\,$AU, and can be adopted as the
outer radius of the dense part of the disc, where multiple scattering
at near-infrared wavelengths is dominant. It is interesting that the
NACO map does not show clear depolarized areas at the same position,
but seems to resolve the transition in vector orientation, while the
larger beam of NICMOS averaged the differently oriented vectors,
resulting in depolarized spots.

The polarized intensity and degree of polarization images in
Fig.~\ref{fig:pol} suggest a slight asymmetry in the dense disc: the
eastern (left) side is straight, while the western (right) side shows
a kink and also has a different position angle than that on the other
side. The thickness of the disc can be measured on the area where the
polarization vectors are aligned (Fig.~\ref{fig:kiskep}). This
approximately corresponds to the area where the degree of polarization
is below ${\approx}\,10\%$. The resulting thickness is approximately
$0.1''$ ($40\,$AU) to the east and is somewhat larger, $0.2''$
($80\,$AU), to the west. One should note that, since these values are
close to the spatial resolution of the polarimetric images, these
numbers should be considered as upper limits for the thickness of the
circumstellar disc around \pars{}. They are upper limits also because
if the inclination is not exactly 90$^{\circ}$, the thickness of the
disc can be even less. The thickness does not show significant
increase with radial distance, suggesting the picture of a flat,
rather than a flared disc, at least considering the dense,
multiple-scattering part. The smooth brightness and polarization
distribution between the disc and the surrounding envelope
(Fig.~\ref{fig:cutok}), however, implies that there is a continuous
density transition between the two components. From the ratio of the
horizontal to vertical sizes of the disc a lower limit of $84^{\circ}$
for the inclination of the system (the angle between the normal of the
disc and the line of sight) can be derived.

\citet{fischer} computed a grid of polarization maps of young stellar
objects with the aim of helping the interpretation of polarimetric
imaging observations. They consider five different models, four with
massive, self-gravitating discs and one with a massless Keplerian
disc. Since the mass of circumstellar material of \pars{} derived from
submillimetre observations is relatively low (${<}\,0.3\,$M$_{\odot}$,
\citealt{henning, polom, sw, hillen}), the most appropriate model for
our case is a Keplerian disc. Indeed the polarization pattern as
computed by \citet{fischer} for an inclination of 87$^{\circ}$ (their
Fig.~1) looks remarkably similar to our Fig.~\ref{fig:vector} (the
differences might be explained by the narrower cavity and flatter disc
of \pars{}). Thus, the geometry and structure assumed by
\citet{fischer} in their Keplerian model could be a good starting
point for further radiative transfer modelling of \pars{}.

\subsection{Modelling the circumstellar environment}
\label{sec:modelling}

\begin{table}
\centering
\caption{Model parameters. Parameters in italics are fixed, while the
  others were fitted.}
\label{tab:par}
\begin{tabular}{lcc}
\hline
Parameter & Variable & Value \\
\hline
Inner disc radius                     & $R_1$         & 3.5\,$R_\odot$ \\
{\it Outer disc radius}               & $R_2$         & {\it 360\,AU}  \\
Temperature at 1 AU                   & $T_{\rm d,0}$ & 285\,K  \\
{\it Power-law index for temperature} & $q_{\rm d}$   & {\it 0.75}  \\
Power-law index for surface density   & $p_{\rm d}$   & 1.6 \\
{\it Disc mass}                       & $M_{\rm d}$   & {\it 0.02\,M$_\odot$} \\
{\it Inclination}                     & $i$           & {\it 86$^\circ$} \\
Inner envelope radius                 & $R_3$         & 5.4\,AU \\
{\it Outer envelope radius}           & $R_4$         & {\it 2000\,AU} \\
Temperature at 5 AU                   & $T_{\rm e,0}$ & 368\,K \\
{\it Power-law index for temperature} & $q_{\rm e}$   & {\it 0.4} \\
Power-law index for surface density   & $p_{\rm e}$   & 0.4 \\
Envelope mass                         & $M_{\rm e}$   & 0.02\,M$_\odot$ \\
{\it Interstellar Extinction}         & $A_{V}$       & {\it 2\,mag}\\
\hline
\end{tabular}
\end{table}

Our observations provide some direct measurements of the geometry of
the circumstellar structure (disc size and thickness, inclination,
envelope size). In the following we discuss the consistency of this
picture with the observed SED. Our approach is to construct a simple
disc+envelope model, in which we fix those parameters whose values are
known from our NACO observations or from other sources (outer disc
radius from this work; power-law index for disc temperature from
\citealt{shakura}; disc mass was set in order to ensure that the whole
disc is optically thick; inclination from this work; outer envelope
radius from this work, the power-law index for envelope temperature is
a typical value for optically thin envelopes containing larger than
interstellar grains, e.g.~\citealt{hartmannkonyv}, Eqn.~4.13;
interstellar extinction from \citealt{hillen}). Then we check whether
the SED can be fitted by tuning the remaining parameters. We adopted
an analytical disk model \citep{adams}, which has been successfully
used to model FUors \citep{quanz2, v1647ori_midi}. Our model consists
of two components, an optically thick and geometrically thin accretion
disc \citep{shakura} and an optically thin envelope (no cavity is
assumed). No central star is included in the simulation, partly
because in outbursting FUors the star's contribution is negligible
compared to that of the inner disc \citep{hk96}, and partly because of
the edge-on geometry where the star is obscured by the disc. This
assumption is supported by the fact that the shape of the SED at
optical wavelengths is broader than a stellar photosphere. The model
also does not take into account internal extinction and light
scattering, thus it cannot reproduce any of the near-IR imaging and
polarimetric observations.

The temperature and surface density distribution in the disc are
described by power-laws:
\begin{equation}
T(r)=T_{{\rm d}, 0}\left(\frac{r}{1\,{\rm AU}}\right)^{-q_{\rm d}}, 
\end{equation}
\begin{equation}
\Sigma(r)=\Sigma_{{\rm d},0}\left(\frac{r}{1\,{\rm
    AU}}\right)^{-p_{\rm d}}.
\end{equation}
Similar power-laws were assumed for the envelope. The observed flux at
a specific frequency is given by
\begin{eqnarray}
F_\nu &=& \frac{\cos{i}}{D^2}\int_{\rm R1}^{\rm R2}2\pi
      r(1-e^{\frac{-\Sigma_{{\rm d}}\kappa_\nu}{\cos{i}}})B_\nu(T_{\rm
      d}){\rm d}r + \\
      & & \frac{1}{D^2}\int_{\rm R3}^{\rm R4}2\pi
      r(1-e^{-\Sigma_{{\rm e}}\kappa_\nu})B_\nu(T_{{\rm e}}){\rm d}r.
\end{eqnarray}
The first term describes the emission of the accretion disc, the
second term describes the radiation of the optically thin
envelope. For the dust opacity we used a constant value of
$\kappa_{\nu}\,{=}$ 1 cm$^2$g$^{-1}$ at $\lambda\,{>}\,1300\,\mu$m,
$\kappa_{\nu}\,{=}\,\kappa_{1300\mu\rm
m}\left(\frac{\lambda}{1300\mu\rm m}\right)^{-1}$ between $1300$ and
$100\,\mu$m and again a constant value of
$\kappa_{\nu}\,{=}\,\kappa(100\,\mu\rm m)$ at
$\lambda\,{<}\,100\,\mu$m.

We fitted the SED via $\chi^2$ minimisation using a genetic
optimization algorithm PIKAIA \citep{charbonneau}. This algorithm
performs the maximization of a user defined function, for which
purpose we used the inverse $\chi^2$. Since there are many photometric
measurements in the mid-infrared domain, but just a few in the
far-infrared, the mid-infrared region has a higher weight during the
fit, compared to the far-infrared domain. Therefore, in order to
ensure an equally good fit at all wavelengths, we divided the SED into
four regions and weighted the $\chi^2$ of each domain with the inverse
of number of photmetric points the region contained. Then the final
$\chi^2$ was the sum of the $\chi^2$ of all regions. The regions we
used were: $0.3-3$, $3-30$, $30-300$ and $300-3000\,\mu$m. The
parameters of the best-fit model, which gives a weighted $\chi^2$ of
0.67, are listed in Table~\ref{tab:par}.  The fitted model SED as well
as the disc and envelope components are overplotted in
Fig.~\ref{fig:sed}.

The model SED is consistent with the observed fluxes. This shows that
the picture of a thin accretion disc and an envelope is consistent
with both the measured SED and the geometry and disc/envelope
parameters inferred from our polarimetric observations. Detailed
modelling of the silicate, PAH, and ice spectral features, as well as
the correct treatment of internal extinction and scattering would
require radiative transfer modelling, which will be the topic of a
subsequent paper.

\subsection{The evolutionary status of \pars{}}
\label{sec:general}

The geometry of FUor models discussed in the literature
(e.g.~\citealt{hk96, tbb}) usually consist of a central star
surrounded by an accretion disc and an infalling envelope with a
wind-driven polar hole. These assumptions are supported by the fact
that they fit well the SED \citep{green, quanz}, the interferometric
visibilities \citep{mg,v1647ori_midi} and the temporal evolution of
the SED \citep{fuors}. In this paper we present the first direct
imaging of these circumstellar structures in a FUor. Our polarimetric
measurements of \pars{} show the existence of a circumstellar disc
which extends from at least 48 to 360 AU. The most striking feature of
the disc is its flatness over the whole observed range. The
short-wavelength part of the SED could be well reproduced using a
radial temperature profile of $r^{-0.75}$
(Sect.~\ref{sec:modelling}). This profile is expected from both a
geometrically thin accretion disc and a flat reprocessing disc.  An
envelope was also seen in the polarization maps of \pars{} and it was
also a necessary component for the SED modelling. Envelopes are
involved in many FUor models and in this paper we present a direct
detection of this model component. Our images reveal that the envelope
can be followed inwards as close to the star as the disc. FUor models
often assume a polar cavity in the envelope, created by a strong
outflow or disc wind. The direct images of \pars{} clearly show the
presence of such a cavity and we also detected a bipolar outflow in
the \pars{} system.

In the recent years, as new interferometric and infrared spectroscopic
observations were published for FUors, the group turned out to be more
inhomogeneous in physical properties than earlier assumed, when mainly
optical photometry and spectroscopy had been available. \citet{quanz}
proposed that some differences might be understood as an evolutionary
sequence. They suggest that FUors constitute the link between embedded
Class I objects and the more evolved Class II objects. Members of the
group exhibiting silicate absorption at $10\,\mu$m are younger and
more embedded (Category 1, e.g.~V346\,Nor); while objects with pure
silicate emission are more evolved (Category 2, e.g.~FU\,Ori and
Bran\,76). There are objects showing a superposition of silicate
absorption and emission, which are probably in an intermediary
evolutionary stage (e.g.~RNO\,1B). \citet{green} also sorted FUors,
based on the ratio of the far-infrared excess and the luminosity of
the central accretion disc, $f_d$ (Equ.~7 in their paper). A large
relative excess ($f_d\,{>}\,5\%$) indicates an envelope of large
covering fraction (V1057\,Cyg and V1515\,Cyg), while low relative
excess means a tenuous or completely missing envelope (Bran\,76 and
FU\,Ori). This is also an evolutionary sequence, as young, more
embedded objects have large envelopes, while around more evolved
stars, the envelope has already dispersed. $f_d$ can also be used to
calculate the opening angle of the envelope, thus a prediction of this
scheme is that the opening angle is becoming wider during the
evolution, probably due to strong outflows during the repeated FUor
outbursts.

The two classification schemes are not inconsistent and one can merge
them into the following evolutionary sequence: (1) the {\it youngest
objects} exhibit silicate absorption and large far-infrared excess
(V346\,Nor, probably also OO\,Ser and L1551\,IRS\,5 belong here); (2)
{\it intermediate-aged objects}, where the silicate feature is already
in emission but there is still a significant far-infrared excess
(V1057\,Cyg, V1515\,Cyg, probably also RNO\,1B and V1647\,Ori); (3)
the most {\it evolved objects} show pure silicate emission and low
far-infrared excess (FU\,Ori, Bran 76). We note, however, that this
classification has some weak points. As \citet{quanz} already
mentioned, an edge-on geometry in a more evolved system may appear as
a younger one. Moreover, during an outburst and the subsequent fading
phase, certain spectral features as well as the global shape of the
SED may change.

\pars{} can be placed in this evolutionary scheme, though one should
keep in mind that because of the nearly edge-on geometry, the
classification of this object is somewhat uncertain. \pars{} displays
silicate emission (Fig.~\ref{fig:pah}). Integrating the flux of the
two components in our simple model (Sect.~\ref{sec:modelling}), and
correcting the apparent disc luminosity for inclination effect
($i=86^{\circ}$, Table~\ref{tab:par}) using Equ.~6 of \citet{green},
we obtained a large relative far infrared excess of
$f_d\,{=}\,75\%$. These two properties place \pars{} into the
intermediate-aged category (though because of its inclination, it may
actually seem younger than it is). Following Equ.~7 of \citet{green},
from the $f_d$ value, we also computed the opening angle of the
envelope. The resulting opening angle of $60^{\circ}$ agrees well with
the angle measured in the direct NACO images (Sect.~\ref{sec:env}).

\pars{} was placed into the evolutionary scheme using two parameters:
the silicate feature and the relative far infrared excess. In the
following we discuss whether its other physical characteristics match
with those of other FUors.
\begin{itemize}
\item[{\it (i)}] Our observations revealed that the circumstellar disc
  of \pars{} is very flat. Due to lack of similar direct measurements
  for other FUors, we can only speculate that perhaps all FUors with
  envelopes have such flat discs. On the other hand, the most evolved
  FUor, FU\,Ori, seems to have no envelope but its disc is probably
  flared \citep{kh91, green, quanz2}. This might suggest that disc
  flaring develops at later stages, when illumination from the central
  source may heat the disc surface more directly.
\item[{\it (ii)}] In a nearly edge-on system like \pars{}, one expects
  to see the $10\,\mu$m silicate feature in absorption. The fact that
  \pars{} has silicate emission indicates that the line of sight
  towards the central region is not completely obscured. Using the
  optical depth of the $15.2\,\mu$m CO$_2$ ice feature, we calculated
  an $A_V\,{=}\,8\,$mag ($A_V\,{=}\, 38.7 A_{15.2 \mu\rm{}m}$,
  \citealt{savage}). This value is surprisingly low compared to
  V1057\,Cyg ($A_V\,{\sim}\,50-100\,$mag, \citealt{kh91}). This
  indicates a much more tenuous envelope, which is also supported by
  the low envelope mass of $0.02\,\rm{}M_{\odot}$ in our modelling.
\item[{\it (iii)}] Following \citet{quanz}, we analysed the profile of
  the $15.2\,\mu$m CO$_2$ ice feature of \pars{}. The inset in
  Fig.~\ref{fig:pah} shows that the feature has a characteristic
  double-peaked sub-structure, very similar to HH\,46\,IRS, an
  embedded young source \citep{boogert}. HH\,46\,IRS is a reference
  case for processed ice. The presence of processed ice in \pars{}
  indicates heating processes and the segregation of CO$_2$ and H$_2$O
  ice, already at this evolutionary stage. Other FUors exhibiting this
  kind of profile are L1551\,IRS\,5, RNO\,1B and RNO\,1C
  \citep{quanz}.
\end{itemize}

The evolutionary state of a young stellar object can also be estimated
following the method proposed by \citet{chen}. According to their
Equ.~(1) we calculated a bolometric temperature of T$_{\rm
bol}=410\,$K for the measured SED. We compared this value with the
distribution of corresponding values among young stellar objects in
the Taurus and $\rho\,$Ophiuchus star forming regions (Chen et
al. 1995). From this check we can conclude that \pars{} seems to be a
class I object, and its age is ${\sim}\,10^5\,$yr. However,
\citet{green} argued that the apparent SED of the disc component
depends on the inclination. Thus we computed T$_{\rm bol}$ also for a
face-on disc configuration and obtained T$_{\rm bol}=1160\,$K,
corresponding to a Class II object. In fact, \pars{} is probably close
to the Class I / Class II border, in accordance with the proposal of
\citet{quanz}.

% ---------------------------------------------------------------------
% SUMMARY
% ---------------------------------------------------------------------

\section{Summary}

We present the first high spatial resolution near-infrared direct and
polarimetric observations of \pars{}, with the VLT/NACO instrument. We
complemented these measurements with archival infrared observations,
such as HST/WFPC2 imaging, HST/NICMOS polarimetry, Spitzer IRAC and
MIPS photometry, Spitzer IRS spectroscopy as well as ISO
photometry. Our main conclusions are the following:

\begin{itemize}
\item[(1)] We argue that \pars{} is probably an FU\,Orionis-type
object;
\item[(2)] \pars{} is not associated with any known rich cluster of
  young stars;
\item[(3)] our measurements reveal a circumstellar envelope, a polar
  cavity and an edge-on disc; the disc seems to be geometrically flat
  and extends from at least 48 to 360 AU from the star;
\item[(4)] the SED is consistent with a simple circumstellar disc +
  envelope model;
\item[(5)] within the framework of an evolutionary sequence of FUors
  proposed by \citet{quanz} and \citet{green}, \pars{} can be
  classified as an intermediate-aged object.
\end{itemize}

% ---------------------------------------------------------------------
% ACKNOWLEDGEMENTS
% ---------------------------------------------------------------------

\section*{Acknowledgments}

We are grateful for the Paranal staff for their support during the
observing run and thank our support astronomers O.~Marco and
N.~Ageorges. For the GFP observations and data reduction we thank
G. M. Williger, G. Hilton and B. Woodgate. Observing time at the
Apache Point Observatory 3.5m telescope was provided by a grant of
Director's Discretionary Time. Apache Point is operated by the
Astrophysical Research Consortium. The Goddard Fabry-Perot is
supported under NASA RTOP 51-188-01-22 to GSFC. CAG is also supported
as part of the Astrophysics Data Program under NASA Contract
NNH06CC28C to Eureka Scientific. This research made use of the SIMBAD
astronomical database. This material is partly based upon work
supported by the National Aeronautics and Space Administration through
the NASA Astrobiology Institute under Cooperative Agreement
No.~CAN-02-OSS-02 issued through the Office of Space Science. The work
was partly supported by the grant OTKA K\,62304 of the Hungarian
Scientific Research Fund.

% ---------------------------------------------------------------------
% REFERENCES
% ---------------------------------------------------------------------

\bibliography{kospal}
\bibliographystyle{mn2e}

% ---------------------------------------------------------------------
% APPENDIX
% ---------------------------------------------------------------------

\appendix

\section{Refined calibration of the Spitzer/IRS beam profiles}

\begin{table}
\caption{Log of IRS calibration measurements.}
\label{tab:calirs}
\centering
\begin{tabular}{c c c c}
\hline
Channel    &   AOR    &   Date      & Target \\
\hline
Short Low  & 16295168 & 2005-Nov-21 & HR 7341 \\
           & 19324160 & 2006-Jul-5  & HR 7341 \\
Short High & 16294912 & 2005-Nov-21 & HR 6688 \\
Long Low   & 16463104 & 2005-Dec-19 & HR 6606 \\
Long High  & 16101888 & 2005-Oct-18 & HR 2491 \\
\hline
\end{tabular}
\end{table}

\begin{figure}
\centering
\includegraphics[angle=0,width=1.0\linewidth]{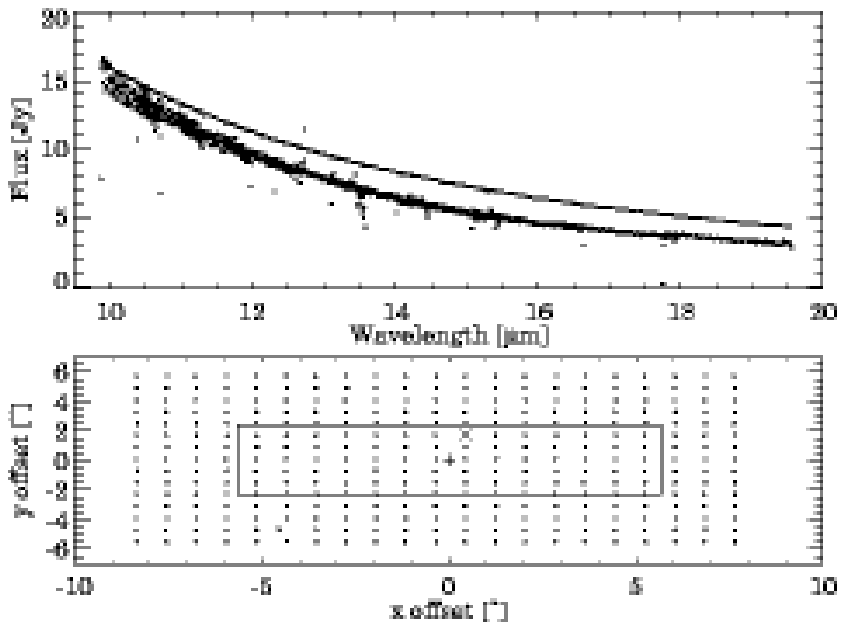}
\caption{{\it Top:} Dots show the measured spectrum of HR\,6688,
obtained with Spitzer/IRS through the Short High slit; the continuous
line show a model spectrum of the same star
\citep[from][]{decin}. {\it Bottom:} Scheme of the data grid; the $+$
sign marks the slit centre; dots show the position of the star
relative to the slit centre; the rectangle indicates the slit itself;
the $\times$ sign marks the position where the spectrum plotted in the
top panel was taken.}
\label{fig:grid}
\end{figure}

\begin{figure}
\centering
\includegraphics[angle=90,width=0.95\linewidth]{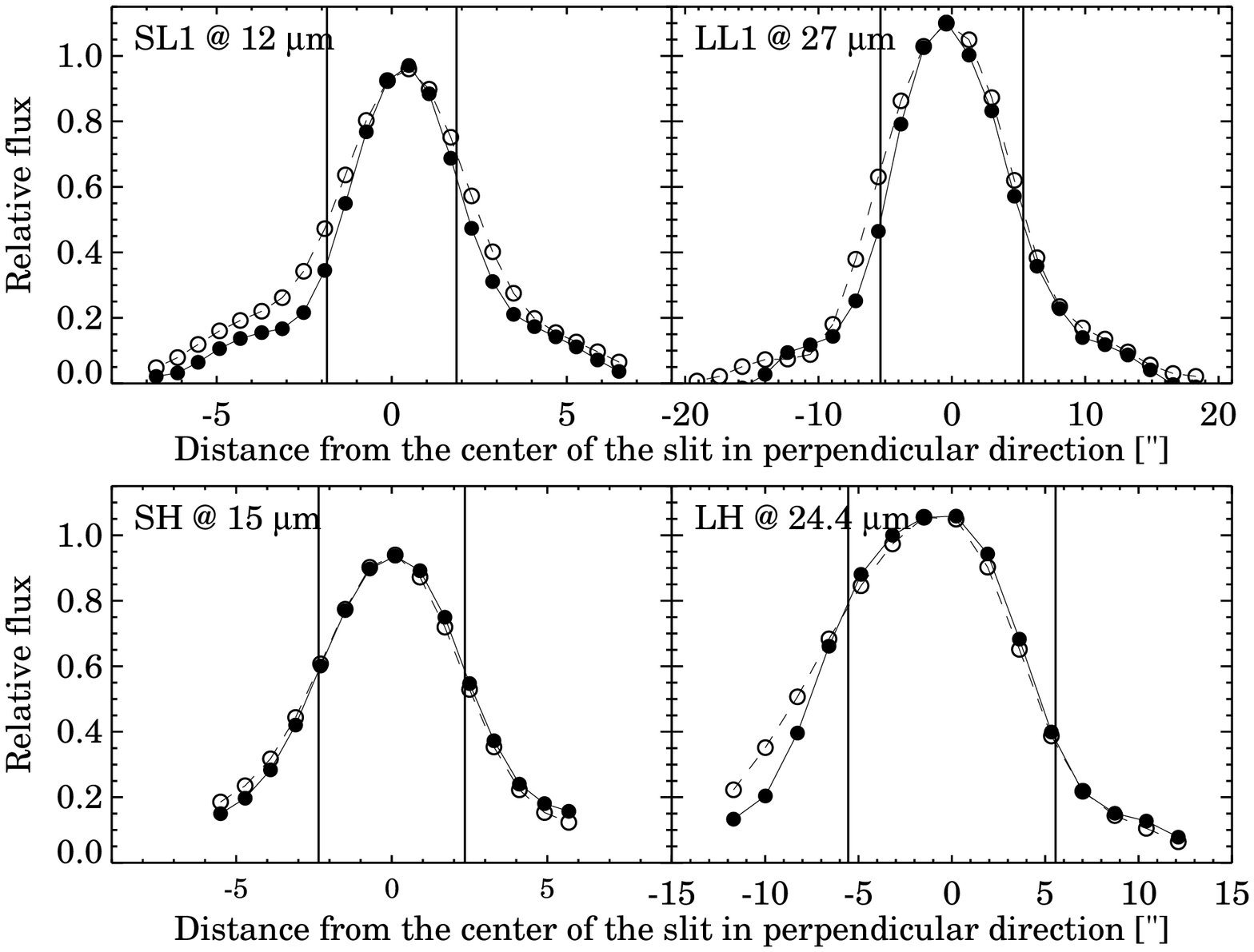}
\caption{Beam profiles of the four Spitzer/IRS channels, perpendicular
  to the slit, at selected wavelengths. Filled dots show the measured
  profile, while open dots indicate the pixelized model PSF.}
\label{fig:IRS}
\end{figure}

In those cases when the source is not well centred in the slit, part
of the stellar PSF falls outside of the slit, leading to flux
loss. When the precise absolute flux level of the spectrum is
important, one should correct for this flux loss. Correction for this
kind of flux loss is not implemented in the Spitzer data reduction
pipeline. For point sources flux loss in off-centred sources can be
efficiently corrected if we know the beam profiles of the different
IRS slits. In order to construct the beam profiles for all four IRS
slits, we reduced and analysed dedicated IRS calibration measurements
taken in spectral mapping mode in a regular
grid. Table~\ref{tab:calirs} shows a log of these calibration
measurements. We used the pipeline processed post-BCD files (pipeline
version S14.0.0). At each spatial position, we divided the measured
spectrum with a synthetic MARCS model spectrum appropriate for the
measured star \citep{decin}. The result is a data cube with two
spatial dimensions and one wavelength dimension. We resampled these
data cubes to finer spatial grid and smoothed them in wavelength.

We checked whether the measured beam profiles are consistent with the
PSFs provided by Spitzer's Tiny Tim (J. Krist). In Fig.~\ref{fig:IRS}
we plotted the ratio between the observed and the model spectrum at
different distances from the slit centre at a certain wavelength. We
found slight differences: the measured profiles were in general
narrower than the model PSFs, and in some cases they were not centred
at zero. We examined several effects that can cause a difference
between the model and the measured profiles and we found that the
differences can be attributed to pixelization effects (pixel sizes are
$1.8''$ for SL, $2.3''$ for SH, $5.1''$ for LL and $4.5''$ for LH) and
to small (few tenths of arcsecond) uncertainties in the exact position
of the calibration star with respect to the slit. Based on our
analysis, we decided to use the measured profiles to correct our
science measurements, and to use the TinyTim model profiles to
estimate the uncertainty of the correction. Using this correction, the
absolute flux level of IRS spectra has an uncertainty of 10\%, but
individual measurements can be much more precise if the source is
well-centred in the slit and the necessary correction is small. More
details on IRS beam profiles will be given in a future paper.

\begin{figure}
\centering
\includegraphics[angle=0,width=0.55\linewidth]{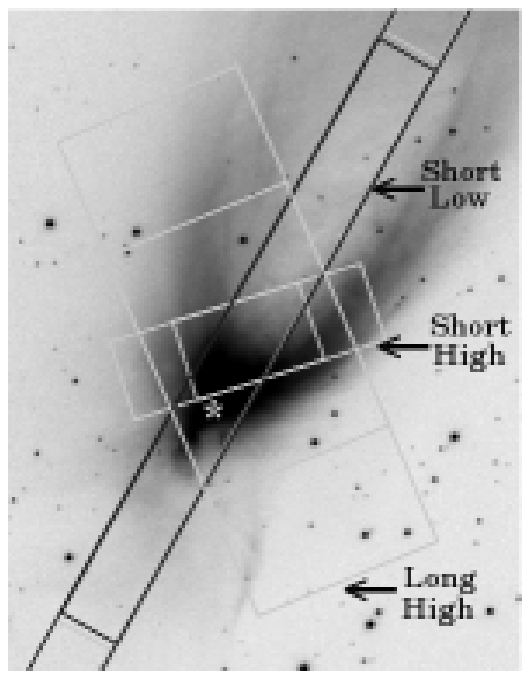}
\caption{Positions of the Spitzer/IRS slits with respect to \pars{}
  (marked with an asterisk). Black rectangle: Short Low, small grey
  rectangle: Short High, big grey rectangle: Long High.}
\label{fig:focal}
\end{figure}

\begin{figure}
\centering
\includegraphics[angle=0,width=0.95\linewidth]{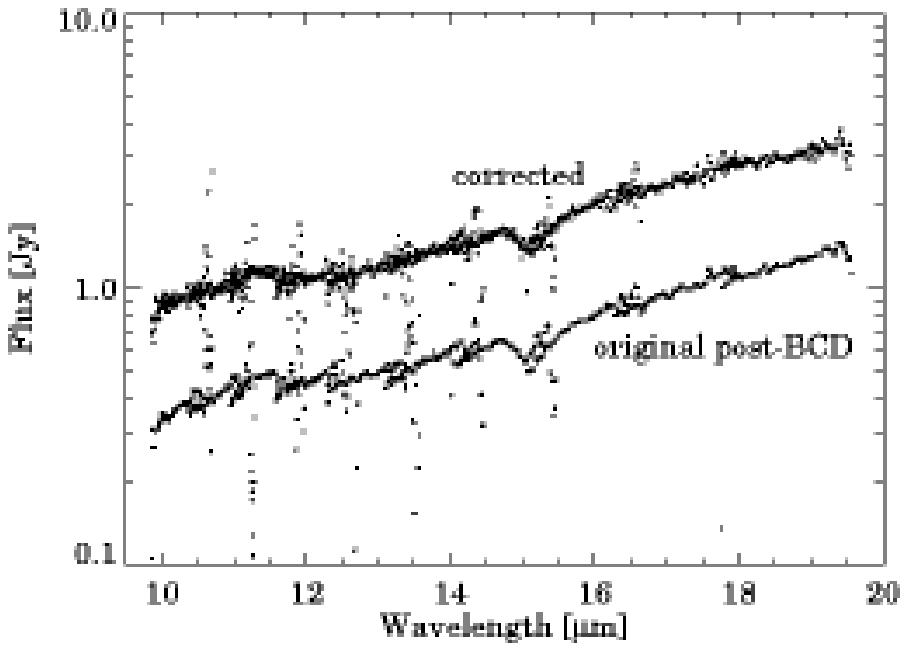}
\caption{Short High channel spectrum of \pars{}.}
\label{fig:sh}
\end{figure}

Figure~\ref{fig:focal} shows the position of the IRS slits overplotted
on the HST/WFPC2 image of \pars{}. While the target is well-centred in
the Short Low slit, it is slightly off-centred in the Long High and it
is practically off-slit in the Short High. Considering our measured
beam profiles, we found that no correction is necessary in the Short
Low channel, a small correction ($5-20\%$, depending on wavelength) is
necessary in the Long High channel, and a significant correction must
be made in the Short High channel, where the corrected flux is
approximately 3-4 times the original one (see
Fig.~\ref{fig:sh}). After the correction, the different IRS channels
match fairly well in the overlapping wavelength regimes.

\label{lastpage}

\end{document}